\documentclass[useAMS,usenatbib]{mn2e}
\usepackage{Times}
\usepackage{psfrag}
\usepackage{pspicture}
\usepackage{graphicx}
\usepackage{amsmath}

\usepackage{array}

\title[The Ly$\alpha$ luminosity function at $\bf z=6.6$]{Identification of the brightest Ly$\alpha$ emitters at $\bf z=6.6$: implications for the evolution of the luminosity function in the re-ionisation era}

\author[J. Matthee et al.]{Jorryt Matthee$^{1}$\thanks{E-mail: matthee@strw.leidenuniv.nl}, David Sobral$^{1,2,3}$, S\'ergio Santos$^{2,3}$, Huub R\"ottgering$^{1}$, \newauthor Behnam Darvish$^{4}$ and Bahram Mobasher$^{4}$\\
$^{1}$ Leiden Observatory, Leiden University, P.O.\ Box 9513, NL-2300 RA Leiden, The Netherlands\\
$^{2}$ Instituto de Astrof\'{\i}sica e Ci\^{e}ncias do Espa\c{c}o, Universidade de Lisboa, OAL, Tapada da Ajuda, PT1349-018 Lisboa, Portugal \\
$^{3}$ Departamento de F\'{i}sica, Faculdade de Ci\^{e}ncias, Universidade de Lisboa, Edif\'{i}cio C8, Campo Grande, PT1749-016 Lisbon, Portugal \\
$^{4}$ University of California, Riverside, 900 University Ave, Riverside, CA, 92521, USA }
\begin{document}

\date{Accepted by MNRAS on 25 April 2015}

\pagerange{\pageref{firstpage}--\pageref{lastpage}} \pubyear{2015}

\maketitle

\label{firstpage}

\begin{abstract}
Using wide field narrow-band surveys, we provide a new measurement of the $z=6.6$ Lyman-$\alpha$ Emitter (LAE) luminosity function (LF), which constraints the bright end for the first time. We use a combination of archival narrow-band NB921 data in UDS and new NB921 measurements in SA22 and COSMOS/UltraVISTA, all observed with the Subaru telescope, with a total area of $\sim 5$ deg$^2$. We exclude lower redshift interlopers by using broad-band optical and near-infrared photometry and also exclude three supernovae with data split over multiple epochs. Combining the UDS and COSMOS samples we find no evolution of the bright end of the Ly$\alpha$ LF between $z=5.7$ and $6.6$, which is supported by spectroscopic follow-up, and conclude that sources with \emph{Himiko}-like luminosity are not as rare as previously thought, with number densities of $\sim 1.5\times10^{-5}$ Mpc$^{-3}$. Combined with our wide-field SA22 measurements, our results indicate a non-Schechter-like bright end of the LF at $z=6.6$ and a different evolution of \emph{observed} faint and bright LAEs, overcoming cosmic variance.
This differential evolution is also seen in the spectroscopic follow-up of UV selected galaxies and is now also confirmed for Ly$\alpha$ emitters, and we argue that it may be an effect of re-ionisation. Using a toy-model, we show that such differential evolution of the LF is expected, since brighter sources are able to ionise their surroundings earlier, such that Ly$\alpha$ photons are able to escape. Our targets are excellent candidates for detailed follow-up studies and provide the possibility to give a unique view on the earliest stages in the formation of galaxies and re-ionisation process.
\end{abstract}
\begin{keywords}
galaxies: high-redshift, galaxies: luminosity function, cosmology: observations, galaxies: evolution, cosmology: dark ages, re-ionisation, first stars.
\end{keywords}

\section{Introduction}
The Lyman-$\alpha$ (Ly$\alpha$) emission line (1216 {\AA}) is a powerful tool to study the formation of galaxies in the early Universe. This is because it has been predicted to be emitted by young $``$primeval$"$ galaxies \citep{PartridgePeebles1967,Pritchet1994}, but also because it is redshifted into optical wavelengths at $z>2$, where most rest-frame optical emission lines are impossible to observe with current instrumentation.

Indeed, the Ly$\alpha$ line has been used to spectroscopically confirm high redshift candidate galaxies up to $z\sim7.5$ obtained with the Lyman-break technique \citep[e.g.][]{Steidel1996,Finkelstein2013,Schenker2014}, which is based on broad-band photometry using e.g. WFC3 on the \emph{Hubble Space Telescope} (HST). Galaxies selected this way are called Lyman-Break Galaxies (LBGs) and the current largest sample contains already 10,000s \citep[e.g.][]{Bouwens2014}. 

Narrow-band surveys select emission line-galaxies at specific redshift slices and are therefore used to search for Ly$\alpha$ emitters (LAEs) directly. Samples of LAEs have now been established from $z\sim2-7$ through narrow-band surveys \citep[e.g.][]{CowieHu1998,Rhoads2000,Fynbo2001,Rhoads2003,MalhotraRhoads2004,Taniguchi2005,Shimasaku2006,Westra2006,Nilsson2007,Ouchi2008,Ouchi2010,Hu2010,Kashikawa2011,Shibuya2012,Konno2014}, but also through spectroscopic surveys  \citep[e.g. HETDEX, VUDS and MUSE;][]{Hill2008,Cassata2015,Bacon2014}. Limited samples of LAEs at lower redshifts and the local Universe also exist that are detected through e.g. GALEX or HST \citep[e.g.][]{Hayes2007,Deharveng2008,Cowie2010}.

While part of the difference between LAEs and LBGs is just the way they are detected, there are also differences in their average properties. There exists an anti-correlation between the UV brightness and the Ly$\alpha$ Equivalent Width (EW) \citep{Ando2006,Stark2010}, indicating that the brightest LBGs are typically not LAEs, and that the UV continuum for most LAEs is very hard to detect, even in the deepest broad-band images \citep{Bacon2014}. Spectroscopic follow-up of LBG selected galaxies has shown that the typical Ly$\alpha$ EW increases with increasing redshift up to $z\sim 6.5$. This is likely due to LBGs being younger on average and less dustier at higher redshift \citep{Stark2010,Schenker2012,Cassata2015}. 
This picture is consistent with the evolution of the luminosity function (LF) of the different classes of galaxies. For LAEs, the Ly$\alpha$ LF is remarkably constant between $z=3-6$ \citep[e.g.][]{Shimasaku2006,Dawson2007,Gronwall2007,Ouchi2008}, while the UV LF of LBGs declines to higher redshifts in a reasonably uniform way due to the decline in the global star formation activity in galaxies \citep{Ellis2013,Bouwens2014,McLeod2014}. This also indicates that the Ly$\alpha$ emission line generally brightens with increasing redshift.

From these observations, the picture has emerged that Ly$\alpha$ is preferentially observed at a specific phase during a galaxy's evolution. Since Ly$\alpha$ is produced by recombination radiation from hydrogen clouds around very massive, young ($<10$ Myr) stars \citep[e.g.][]{Schaerer2003}, and Ly$\alpha$ is easily absorbed and re-scattered (leading to lower surface brightnesses), on average LAEs are believed to be young starbursts, while LBGs on average are slightly more evolved galaxies with a higher dust content \citep[e.g.][]{Verhamme2008}. \cite{Ono2010} find that the UV slope of $z=6-7$ LAEs is very steep ($\beta = -3$), while \cite{Bouwens2014slope} find that the UV slope of LBGs at similar redshifts is typically slightly shallower ($\beta = -2.2$). From clustering measurements, \cite{Gawiser2007,Ouchi2010} and \cite{Bielby2015} agree on an average LAE halo mass of $\sim10^{11}$ M$_{\odot}$ from $z=3-7$. For LBGs alternatively, the typical halo mass is one order of magnitude higher at these redshifts \citep[e.g.][]{Ouchi2003,Hamana2004,Ouchi2005,Hildebrandt2009}, more typical of $``$Milky Way$"$ dark matter haloes of $10^{12}$  M$_{\odot}$.

Near the re-ionisation redshift, physical processes start to play a role which are additional to intrinsic changes in the properties of galaxies, since Ly$\alpha$ is easily absorbed by a neutral Inter Galactic Medium (IGM). While LAEs can be an important source of ionising photons for re-ionisation, one of the main interests in studying LAEs at this epoch is observable effects of a higher neutral IGM opacity. Besides evolution of the luminosity function, these observables include an increased observed clustering in a more neutral IGM, and attenuated line-profile. The observed clustering of LAEs increases since the observability is favoured when sources are in overlapping ionised spheres \citep{McQuinn2007}, while the line-profile becomes more asymmetric due to absorption and re-scattering of Ly$\alpha$ photons in a more neutral medium \citep{Dijkstra2007}. 

At $z>7$ spectroscopic follow-up of LBGs is remarkably less successful \citep{Fontana2010,Stark2010,Pentericci2011,Ono2012,Treu2013,Pentericci2014,Caruana2014}, indicating either a lower intrinsic escape of Ly$\alpha$ \citep[e.g.][]{Dijkstra2014}, an increased column density of absorbing clouds \citep{BoltonHaehnelt2013}, or a higher neutral fraction of the IGM \citep{Santos2004,Dijkstra2007,Schenker2014,Taylor2014,Tilvi2014}. There is also evidence for an increased opacity to Ly$\alpha$ photons from the Ly$\alpha$ LF, which is observed to decline very rapidly \citep{Ouchi2010,Konno2014}. Searches for LAEs at $z=7.7$ and $z=8.8$ have been unsuccessful in spectroscopically confirming any of the candidates \citep{Willis2005,Cuby2007,Willis2008,Sobral2009,Hibon2010,Tilvi2010,Clement2011,Krug2012,Jiang2013,Faisst2014,Matthee2014}. However, these studies are still limited by their sensitivity since Ly$\alpha$ is shifted into the near-infrared (NIR). Most of these studies only probe tiny areas in the sky, meaning that bright sources might be missed. 

Typically, research is so far limited to $\sim1$ deg$^2$ areas \citep[e.g.][]{Ouchi2008,Ouchi2010}, where cosmic variance, especially for the observability of Ly$\alpha$ around the re-ionisation epoch, can play a large role. To make further progress, we are carrying out an extensive set of wide-field narrow-band observations to study the evolution of Ly$\alpha$ emitters from the epoch of re-ionisation ($z\sim6-9$) up to the peak of the cosmic star formation history ($z\sim2-3$). Our aim is to explore the evolution of the bright end which is so far uncharted and for which spectroscopic follow-up is easier and gives a better comparison to to surveys at the highest redshifts. \\

In this paper, we focus at the $z=6.6$ Ly$\alpha$ LF because of its importance to the study of re-ionisation. The widest narrow-band survey at that redshift to date has been presented by \cite{Ouchi2010}, which reaches a Ly$\alpha$ luminosity of $\sim 10^{42.5}$erg s$^{-1}$ over a $\sim0.9$ deg$^2$ area. The brightest Ly$\alpha$ emitter in their sample, \emph{Himiko}, with a luminosity of $3.5\times10^{43}$ erg s$^{-1}$\citep{Ouchi2009,Ouchi2013}, has been seen as a very rare source, a triple merger, one of its kind. Because there has been only one very bright source known, the error on its number density is large, such that there is a factor 30 offset between the fitted luminosity function and the data. We have obtained wide field observations to further constrain the number density of bright Ly$\alpha$ emitters. We introduce our data and sample selection in \S 2. Using our data set, we re-produce the \cite{Ouchi2010} sample and add new $z=6.6$ Ly$\alpha$ candidates using deep archival Subaru data in the COSMOS field and new shallow wide-field data in SA22 in \S 3. Our estimates of the completeness of our selection procedure and description of the corrections made to the luminosity function are shown in \S 4. This leads to a new estimate of the luminosity function in \S 5 (supported by spectroscopic confirmation of the two brightest LAEs in COSMOS; \cite{SobralCR7}), where we find that the combined luminosity function has a non Schechter-like bright end. We discuss the evolution and implication for re-ionisation in \S 6. The main results are shown in Fig. $\ref{fig:lf_final}$, which shows our new estimate of the $z=6.6$ Ly$\alpha$ LF, and Fig. $\ref{fig:lf_evo}$, where we compare the evolution between $z=5.7$ and $z=6.6$. \\
Throughout the paper, we use a 737 $\Lambda$CDM cosmology (H$_0$ = 70 km s$^{-1} $Mpc$^{-1}$, $\Omega_M = 0.3$, $\Omega_{\Lambda} = 0.7$) and magnitudes are measured in 2" diameter apertures in the AB system.

\section{Observations \& Data reduction}
\subsection{Imaging}
\begin{figure}
\centering
\includegraphics[width=8cm]{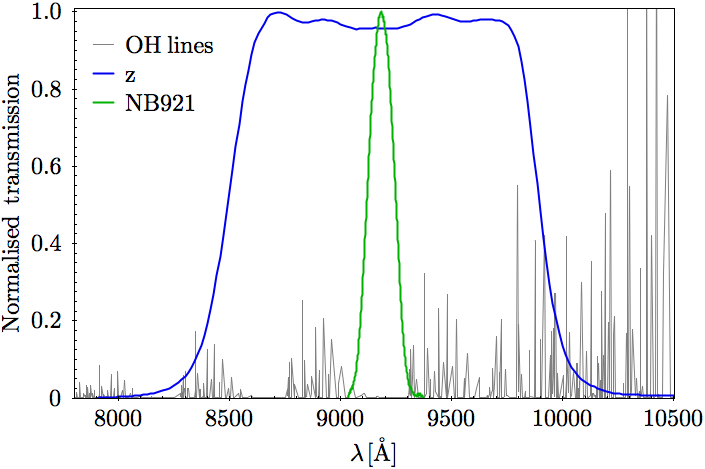}
\caption{\small{Filter transmission profile of the NB921(blue) and $z$ (green). The transmission is normalised to the maximum transmission in each filter. In grey, OH emission lines from the night sky are shown \citep{Rousselot2000}. It can be seen that the narrow-band is in sky-line free region allowing us to go very deep. This also facilitates the spectroscopic follow-up.
}}
\label{fig:filter}
\end{figure}

Optical imaging data was obtained with Subaru's Suprime-Cam \citep{Miyazaki2002} using the NB921 narrow-band filter (Fig. $\ref{fig:filter}$). The Suprime-Cam is composed by 10 CCDs with a combined field of view of $34' \times 27'$ and with chip gaps of $\sim15''$. The NB921 filter has a central wavelength of 9196 \AA\  and a FWHM of 132 {\AA} and is located in a wavelength region free of OH line-emission in the atmosphere (see Fig. $\ref{fig:filter}$).\\ 

\begin{table*}
\label{table:obs}
	\begin{center}
		\label{tab:observationlog}
		\begin{tabular}{@{}cccccccc@{}}
		\hline
		$\bf Field$ & $\bf R.A.$ & $\bf Dec.$ &  $\bf Int. time$ & $\bf FWHM$ & $\bf Area$ & $\bf Depth$ & $\bf Dates$\\
 		 & (J2000) & (J2000) & (ks) & ('') & (deg$^2$) & ($3\sigma$) &  \\
		\hline
		COSMOS-1  & 10 01 28 & +02 25 51 & 10.8 & 0.3 & 0.24 & 25.8 & 2009 Dec 19 \\
		COSMOS-2  & 09 59 35 & +02 27 01 & 8.78 & 0.3 & 0.24 & 25.9 & 2009 Dec 19,20\\
		COSMOS-3  & 10 01 24 & +01 58 00  & 10.8 & 0.6 & 0.24 & 25.9 &2009 Dec 21\\
		COSMOS-4  & 09 59 29 & +01 58 42 & 7.80 & 0.6 & 0.24 & 25.7 & 2009 Dec 21 \\
		SA22-DEEP  & 22 19 14 & +00 11 24 & 32.1 & 0.3 & 0.17 & 26.4 & 2009 Sep 15-17\\
		SA22-WIDE-[1-19]{\bf *} & 22 16 19 & +00 10 00 & 0.36 & 0.5 & 2.72 & 24.3 & 2014 May 28-29\\
		UKIDSS-UDS C  & 02 18 00 & -05 00 00 & 30.0 & 0.8  & 0.14 & 26.4 &  2005 Oct 29, Nov 1, 2007 Oct 11,12\\
		UKIDSS-UDS N  & 02 18 00 & -04 35 00  & 37.8 & 0.9 & 0.18 & 26.4 &  2005 Oct 30,31, Nov 1, 2006 Nov 18, 2007 Oct 11,12\\
		UKIDSS-UDS S  & 02 18 00 & -05 25 00 & 37.1 & 0.8& 0.19 & 26.4 &  2005 Aug 29, Oct 29, 2006 Nov 18, 2007 Oct 12\\
		UKIDSS-UDS E  & 02 19 47 & -05 00 00  & 29.3 & 0.8 & 0.16 & 26.4 & 2005 Oct 31, Nov 1, 2006 Nov 18, 2007 Oct 11,12\\
		UKIDSS-UDS W & 02 16 13 & -05 00 00 & 28.1 & 0.8 & 0.14 & 26.4 &  2006 Nov 18, 2007 Oct 11,12 \\
		\hline
\end{tabular}
\end{center}
	\caption{\small{Observation log for the NB921 optical imaging in the COSMOS, SA22 and UKIDSS UDS fields. Depths are based on empty aperture measurements and take correlations in the background into account. The UDS data have been analysed by \citet{Ouchi2010}. The area is the area after masking (see \S 2.1). \textbf{*} Note that we have 19 pointings in SA22-WIDE, all with similar observing conditions.}}
\end{table*} 

We obtained archival ultra-deep observations in UKIDSS-UDS  (02 18 00 -05 00 00) and COSMOS-UltraVISTA (10 00 00 +02 00 00)  and SA22 (22 00 00 +00 00 00) and took new data in a wide area ($\sim4.5$ deg$^2$) in SA22 on May 28-29 2014, observing program S14A-086 (PI: Sobral). These fields were chosen for their multi-wavelength coverage and low Galactic foreground emission. Apart from providing a large total area, the combination of ultra-deep and shallower surveys allows us to sample a wide range of luminosities.\\ 
In UDS, there are 5 sub-fields with a total integration time ranging from 7.8 to 10.5 hours (see Table 1). This exposure time is obtained after stacking individual exposures of 1.2ks, with a small dithering pattern. The seeing FWHM ranges from $0.8-0.9"$. We mask regions around bright stars, where spherical artefacts boost the fluxes artificially. We also mask horizontal and vertical stripes caused by blooming of a saturated bright object (see Fig. $\ref{fig:radec}$). This is the same raw data as used to study Ly$\alpha$ emitters at $z=6.6$ by \cite{Ouchi2010}. \\
In the four pointings in COSMOS, the total integration time ranges from 2.2 to 3 hours, with individual exposure time of 1.2ks, such that the total number of exposures per field is smaller (ranging from 7 to 9). In two pointings, the seeing is particularly good ($0.3"$), the other two have seeing FWHM of $0.6"$. Similar to in UDS, we mask spherical halo-regions around bright stars, and vertical and horizontal blooming-stripes (depending on the position angle rotation of the Suprime-cam pointing).\\
The deep SA22 data consists of 27 exposures of 1.2ks at a single pointing with small dithering in perfect seeing conditions (FWHM 0.3"). We mask a small noisy region and blooming patterns. The wide SA22 data consists of 19 pointings of 3 exposures of 120s in very good seeing conditions (FWHM 0.5"). Because of the limited number of exposures, a significant area has a lower signal to noise due to the dithering pattern. Another minor issue is that the astrometric corrections and calibration of the zero-point are less accurate in certain detectors due to the low signal to noise. We conservatively mask all these regions and also mask regions around bright stars and are left with a final coverage of $2.7$ deg$^2$.\\ 
The total area after masking is 4.66 deg$^2$. An overview of the observations is given in Table 1.

\subsection{Data reduction}
The NB921 imaging data was reduced with SDFRED2 package \citep{Ouchi2004RED}. The sequential steps in the reduction are:
\begin{enumerate}
\item \textbf{Overscan and bias subtraction:} for each image, a median value for the overscan region was determined and subtracted in each line of pixels. The bias was subtracted by assuming that it has the same value as the overscan.
\item \textbf{Flat fielding:} flat frames are obtained by observing a uniform light source and are required to correct variations in pixel-to-pixel sensitivity across the camera. By dividing the images by these flat frames, the background becomes flat and luminous patterns caused by differences in the sensibility of the pixels are removed.
\item \textbf{Point spread function homogenisation:} the point spread function (PSF) measures the response of the detector to a point-like source. PSF sizes were obtained by measuring the FWHM of the point-like sources of each frame. The target PSF for homogenisation was defined as the one with more occurrences and the frames with PSF smaller than the target were smoothed with a gaussian.
\item \textbf{Sky background subtraction:} a mesh pattern was computed to represent the sky background, the pattern was interpolated and subtracted on each frame.
\item \textbf{Bad pixel masking:} defects with the detector and problems with the observations may cause data in some pixels to become corrupted. A mask is applied to these pixels.
\item \textbf{Astrometric calibrations}: we correct each image for astrometric distortions using Scamp \citep{Bertin2006}, which fits a polynomial solution by matching detected sources with the 2MASS catalog in the $J$ band \citep{Skrutskie2006}. It also takes into account that images have different integration times and attributes a different weight to each image.
\item \textbf{Stacking}: Once each individual frame has been reduced, we stack the different jittered frames for each pointing.
\item \textbf{Cosmic ray rejection:} cosmic rays are rejected automatically based on the standard deviation in the pixel values in a $1"$ aperture. The standard deviation is typically a hundred times higher for cosmic rays than for real sources. We use a very conservative cut since we do not want to risk rejecting real sources, meaning that our sample will still be somewhat contaminated by cosmic rays. Partly because of this, we inspect all our final candidates visually.
\end{enumerate}

\subsection{Photometric calibration \& survey depths}
Once we obtained the reduced data for each pointing, we set the zero-point (ZP) to a magnitude of 30 (AB). This is done by extracting sources with {\sc SExtractor} \citep{Bertin1996} with high detection thresholds ($>20\sigma$) and match these sources with public catalogues in UDS \citep{Cirasuolo2007}, COSMOS \citep{Ilbert2009} and our own catalog in SA22 \citep[based on $K$ detected sources in UKIDSS-DXS;][]{Matthee2014,SobralCFH},  using {\sc STILTS} \citep{Taylor2006}. We only include sources with narrow-band magnitudes brighter than 19, such that our detections are at sufficient high signal to noise, and fainter than 16, since brighter sources are saturated in our data. In each pointing, we use roughly 500 sources. We then set the ZP by correcting the cropped mean difference between the magnitudes in our images and the ones in the catalogue.\\

We estimate our survey depth by measuring the root mean square (rms) of the background in a million empty apertures with $2"$ diameters, placed at random places in the image, avoiding sources which are detected at $>3\sigma$. Empty aperture measurements take into account that the background noise is correlated and are thus more robust than if the background is measured on a pixel by pixel basis \citep[c.f.][]{MilvangJensen2013}. Also, empty apertures can still include very faint sources below our detection threshold, so this is a conservative upper limit. The depths of the narrow-band images are listed in Table 1. The 3$\sigma$ depths are 26.4 in UDS, 25.8 in COSMOS, 26.4 in the deep pointing in SA22 and 24.3 in the wide pointings in SA22.

\begin{table*}
\label{table:data}
	\begin{center}
		\label{tab:bbdepths}
		\begin{tabular}{@{}cccccc@{}}
		\hline
		$\bf Field$ & $\bf Surveys$ & $\bf Optical$ & $\bf Depths $ &  $\bf Near-infrared$ & $\bf Depths$\\
		\hline
				UDS & SXDF, UDS \& SpUDS & $BVRiz$ & 28.3, 28.4, 27.8, 27.2, 26.6 & $JHK$ & 25.7, 24.5, 24.4 \\
		COSMOS  & COSMOS, UltraVISTA \& S-COSMOS & $BViz$ & 27.6, 27.0, 26.9, 25.8 & $YJHK$ & 25.9, 25.4, 25.1, 24.7 \\
		SA22 & CFHTLS \& DXS & $ugriz$ & 26.2, 26.3, 26.0, 25.7, 24.5 & $JK$ & 24.4, 23.9\\
		\hline
\end{tabular}
	\caption{\small{3$\sigma$ depths of broad-band coverage of our survey-fields obtained using empty aperture measurements. Note that there can be more wavelength data available, but we only use the broad-bands listed here for consistency between the fields. We obtain our own photometry by first registering all the images to the Subaru NB921 measurements and then extracting photometry within 2$"$ apertures in dual mode. The SXDF data is presented in \citet{Furusawa2008}, the COSMOS data in \citet{Ilbert2009} and UltraVISTA in \citet{McCracken2012}.}}
\end{center}
\end{table*} 

\subsection{Multi-wavelength data and photometry}
For UDS, deep $z$-band data (26.6, 3$\sigma$) is available from the Subaru Extreme Deep Field (SXDF) project \citep{Furusawa2008}, as well as data in the optical bands $B$, $V$, $R$ and $i$, with 3$\sigma$ limits: 28.3, 28.4, 27.8 and 27.2 respectively (see Table 2.). This multi-wavelength data is essential to identify different line-emitters. The images in the optical and NB921 are all aligned, because all are from a single survey, telescope and instrument. Furthermore, UKIDSS NIR $J$, $H$, $K$ data \citep{Lawrence2007} is available for 60 \% of the coverage, with 3$\sigma$ depths 25.4, 24.7 and 24.9. For NIR photometry, we use {\sc Swarp} \citep{Bertin2010} to align the NIR images to the NB921 images and interpolate the NIR images, since the pixel scale is slightly larger \citep[UKIRT WFCAM;][]{Casali2007}  than the Suprime-Cam pixel scale.\\

The COSMOS field is one of the best studied extra-galactic fields with $>30$ bands coverage \citep{Ilbert2009}, ranging from X-ray to radio. We use deep optical $BViz$ data from Subaru imaging, with 3$\sigma$ depths of 27.6, 27.0, 26.9 and 25.8 (Table 2) which is available through the COSMOS archive\footnote{\cite{Capak2007}; http://irsa.ipac.caltech.edu/data/COSMOS/}. We align the optical images to the narrow-band images using {\sc Swarp}. NIR data in $YJHK_s$ is available from UltraVISTA DR2 \citep{McCracken2012} with 5$\sigma$ depths 25.4, 25.1, 24.7 and 24.8. The pixel scale of VISTA's VIRCAM is 0.15"/px, so we degrade the images to the pixel scale of the narrow-band images (0.2"/px) and align them. \\

The SA22 field is covered by CFHTLS and UKIDSS DXS surveys. SA22 is W4 in CFHTLS\footnote{http://www.cfht.hawaii.edu/Science/CFHTLS/} and is imaged in $ugriz$ with MegaCam, which has a field of view of 1 x 1 deg$^2$ and a pixel scale of 0.187"/pixel. The near-infrared UKIDSS DXS survey has imaged it in $J$ and $K$ filters with UKIRT/WFCAM.  Note that the multi-wavelength data is not as deep as in the other two fields (typically 1-2 magnitudes shallower), and there is also no Spitzer/IRAC data available. This limits the search for LAEs in the deep pointing since the uncertainty in the $z$-band is much higher than the uncertainty in NB921. For the brightest objects (including all reliable detections in the Wide coverage), this is less of a problem. As before, we align the optical and near-infrared images to our narrow-band pointings using {\sc Swarp}, including degrading the pixel scale to that of the narrow-band imaging.\\

For all fields, photometry is extracted using {\sc SExtractor} in dual-mode with the NB921 image as detection image and within a 2$"$ circular aperture. In the case of a non-detection in any of the broad-bands, we assign 1$\sigma$ limits. We then use the $i$-band to correct our $z$-band such that the median narrow-band excess is zero for all sources by fitting a linear relation between the $(i-z)$ colour and the narrow-band excess. Since the narrow-band filter is almost in the center of the broad-band filter, the correction is small, $z_{cor} = z - 0.13 (i-z) +0.286$. For sources undetected in the $i$ band we assign the median correction of $+0.03$.\\

\section{Selecting Lyman-$\alpha$ emitters at $\bf z=6.6$}
In the NB921 data, Lyman-$\alpha$ emitters need to be selected as line-emitters at $z=6.55\pm 0.055$. This requires multi-wavelength coverage of the fields, which is available through a combination of large (public) surveys. First of all, we require broad-band photometry over the same wavelength coverage as the narrow-band, which is in this case the $z$-band. \\

Line-emitters are selected based on two criteria \citep[e.g.][]{Sobral2013}: the first is that the narrow-band excess must be high enough. Since the observed equivalent width of emission lines increases with redshift and the intrinsic EW of Ly$\alpha$ is high ($\sim 100-200$ {\AA}), we expect Lyman-$\alpha$ emitters to have a high narrow-band excess. We follow previous searches \citep[e.g.][]{Ouchi2010} and use an excess criterion of $z-NB921 > 1$, corresponding to a $z=6.6$ rest-frame EW of $38$ {\AA}. This limit is also chosen to minimise contamination by lower redshift interlopers, although we will also lower this criterion and comment on the differences. To convert the narrow-band excess to the observed EW, we first transform magnitudes ($m_i$) to flux densities in each filter ($f_i$) with the standard AB convention:

\begin{equation}
f_i = \frac{c}{\lambda^2_{i, center}} 10^{-0.4(m_i +48.6)}
\end{equation}

In this equation, $c$ is the speed of light and $\lambda_{i, center}$ is the central wavelength in each filter, which are 9183.8 {\AA} and 8781.7 {\AA} for the narrow-band ($NB$) and broad-band ($BB$) respectively.\\
Using Eq. 1, we use the following equations to convert to EW and line-flux respectively:
\begin{equation}
EW = \Delta\lambda_{NB} \frac{f_{NB}-f_{BB}}{f_{BB}-f_{NB} \frac{\Delta\lambda_{NB}}{\Delta\lambda_{BB}}}
\end{equation}

Here, $f_{NB}$ and $f_{BB}$ are the flux-densities, $\Delta\lambda_{NB}$ and $\Delta\lambda_{BB}$ the filter-widths, 135.1 {\AA} and 1124.6 {\AA}, respectively. In this formula, the numerator is the difference in narrow-band and broad-band flux and the denominator the continuum, which is corrected for the contribution from the narrow-band flux. The formula breaks down at a certain $NB$ excess depending on the specific filters, and thus we set the EW of those sources to $>1500$ {\AA}. \\
The line-flux is computed using:
\begin{equation}
f_{line}= \Delta\lambda_{NB} \frac{f_{NB}-f_{BB}}{1-\frac{\Delta\lambda_{NB}}{\Delta\lambda_{BB}}}
\end{equation}

The second criterion for selecting emission line galaxies is that the excess should be significant, meaning that it is not dominated by errors in the narrow-band and broad-band photometry. We will follow the methodology presented in \cite{Bunker1995} and the equation from \cite{Sobral2013} to compute the excess significance ($\Sigma$):

\begin{equation}
\Sigma = \frac{1-10^{-0.4(BB-NB)}}{10^{-0.4(ZP-NB)}\sqrt{\pi r_{ap}^2(\sigma^2_{px,BB}+\sigma^2_{px,NB})}}
\end{equation}
In this case, $BB$ is the $z$-band magnitude after correction using the i band (see next subsections), $NB$ the magnitude in NB921, $ZP$ the zero-point of the images, which is set to a 30 AB magnitude. $\sigma_{px}$ is the root mean squared (rms) of background pixel values in the data of the respective filters and $r_{ap}$ is the aperture radius in pixels. Our depths are estimated using empty aperture based rms values, which takes correlations in the background into account. For the selection of emitters however, we use pixel based rms values for consistency with previous surveys, but also check that our results are robust when using empty aperture-based $\Sigma$ values. \\

After selecting a sample of line-emitters, we use multi-wavelength data to distinguish high redshift candidate LAEs. In addition to the $z$-band, we also need bands in bluer wavelengths in order to apply the Lyman break technique to select high redshift sources. In this case this means that there should be no detection in the $B$, $V$, $u$, $g$ and $r$ filters and a strong break in the $(i-z)$ colours. Measurements in redder wavelengths, such as in the near-infrared (NIR) $J$, $H$ and $K$ filters, can provide valuable insight in the nature of the candidates and possibly help excluding lower-redshift interlopers. Finally, Spitzer-IRAC data can be used as further constraints on excluding dusty low redshift interlopers (generally with bright IRAC detections and red colours), or as a further evidence for the source being at $z=6.6$, since at that redshift the $[3.6]$ and $[4.5]$ $\mu$m bands are contaminated in such a way that sources with strong nebular emission (EW $>1000$ {\AA}) are expected to have blue $[3.6]-[4.5]$ colours \citep[e.g.][]{Stark2013,Smit2014,Smit2015}. \\
In addition to using colours for our selection and characterisation of Ly$\alpha$ candidates, we also compute photometric redshifts using EAZY v1.1 \citep{Brammer2008}, which includes the contribution of emission lines (although typically not strong enough for the observed extreme emission lines galaxies). We include optical and near-infrared photometry and we use it to identify possible lower redshift interlopers.\\
The details of our selection per field are presented in \S 3.1 for UDS, \S 3.2 for COSMOS and \S 3.3 for SA22. 

\begin{figure*}
\centering
\includegraphics[width=15cm]{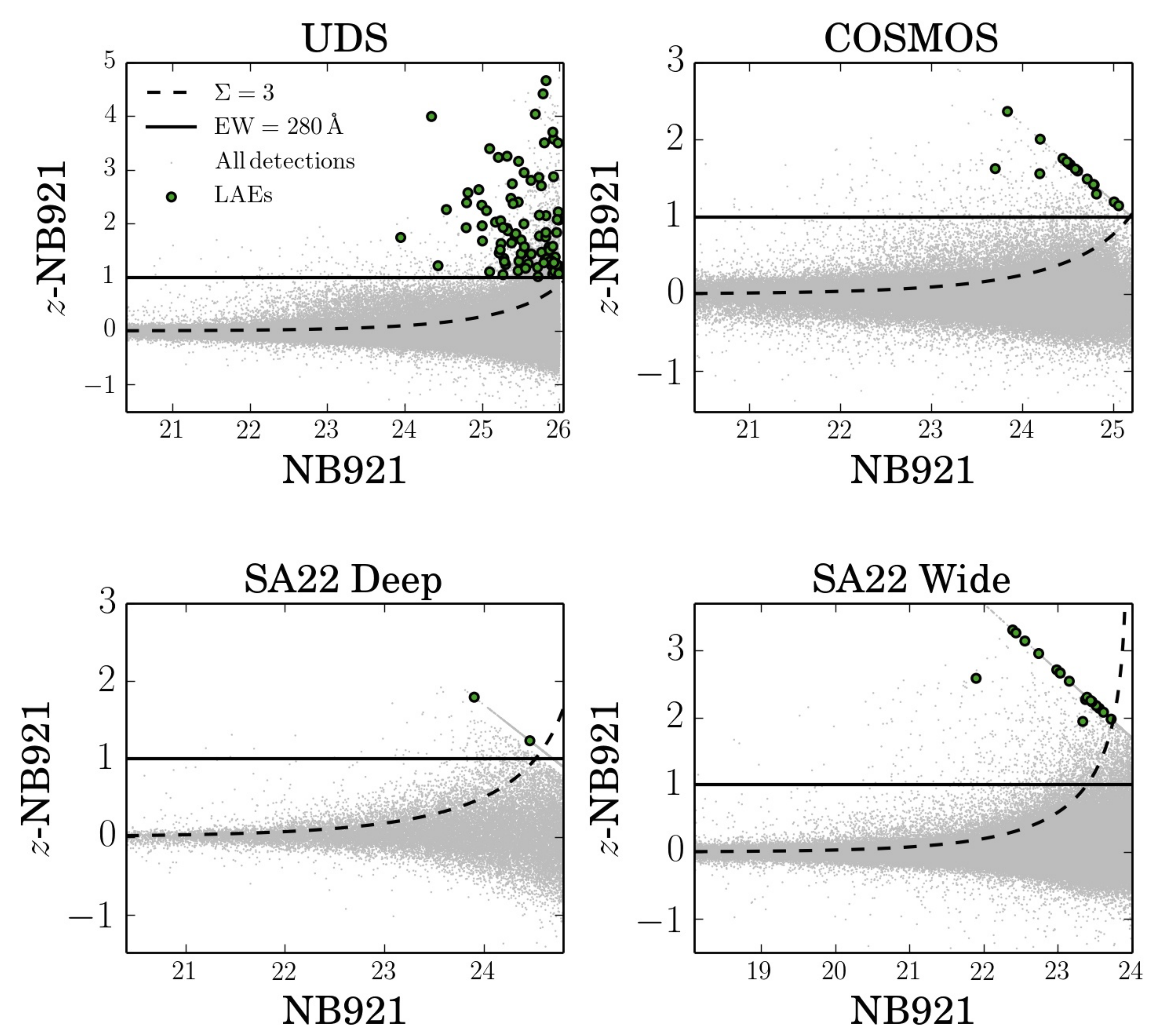}
\caption{\small{Narrow-band excess versus narrow-band magnitude. This figure shows the selection of line-emitters in our different fields. The grey points show all detected objects (after removing cosmic rays, visually identified spurious sources and sources for which the excess is unphysical) and the green points show the selected LAEs in each field. The selection consists of an EW cut (solid horizontal line) and a criterion to determine whether the excess is significant (dashed lines). The EW cut corresponds to a restframe EW of 38 {\AA} at $z=6.56$, and the significance cut corresponds to 3$\Sigma$. After visual removal of spurious sources and cosmic rays and optical and near-infrared broadband criteria, we find 99 LAEs in UDS, 15 in COSMOS, 2 in SA22-Deep and 19 in SA22-Wide. While the narrow-band imaging in UDS and SA22-Deep have similar depth, the number in UDS is higher due to a roughly 5 times larger area and (most importantly) deeper $z$-band imaging. There are less candidates in COSMOS due to shallower narrow-band imaging. The imaging in SA22-Wide is even shallower, but this is compensated by a much larger area. }}
\label{fig:excess}
\end{figure*}  

\subsection{Selecting LAEs in UDS}
We use our photometry described in \S 2.4 and select line-emitters using the following criteria: \\ 
\begin{enumerate}
\item  $\bf z-NB921 > 1$\\
\item ${\bf \Sigma > 3}$\\ 
\item Pass visual inspection\\
\end{enumerate}
After the first two criteria, we have 1514 line-emitter candidates (see Table 3 for the numbers after each step). We check each line-emitter candidate visually in the NB921 image and exclude 122. Furthermore, 25 candidates are excluded since their excess is unphysically high, meaning that their (non-)detection in the $z$-band is in disagreement with the flux solely contributed by the measured narrow-band flux by more than 3$\sigma$. Most of the excluded candidates have their flux boosted by artifacts from bright stars, such as haloes or spikes (even after masking), others are excluded because they reflect read-out noise (since the images are so deep) and there are CCD-grid like patterns. 

In total, we find 1367 line-emitters, which selection is shown in Fig. $\ref{fig:excess}$. This sample is dominated by H$\beta$/{\sc [OIII]} at $z=0.83$, {\sc[OII]} emitters at $z=1.46$ and H$\alpha$ at $z=0.40$ (based on photometric redshifts), even though the high excess criterion is already used to minimise this number. These lower redshift emitters are described for example in \cite{Sobral2013}, which also shows a photometric redshift distribution. To select Ly$\alpha$ candidates, we exclude low redshift sources and select high-redshift candidates using the Lyman-break technique:
\begin{multline}  \nonumber B > 28.7 \land V > 28.2  \land ( i-z > 1.3 \, \, {\vee} \,\, i > 27.2 )   
\end{multline}
The Lyman-break technique is based on the absence of flux blue-ward of the Lyman limit (912 {\AA}) and we therefore require our sources not to be detected at wavelengths below $912 \times (6.6 + 1) = 6930$ {\AA}. This results in non-detections in $B$ and $V$, and a strong $(i-z)$ break (or a non-detection in $i$ as well). The $(i-z)$ criterion, adapted from \cite{Ouchi2010} takes the Gunn-Peterson trough at $z=6.6$ into account. For consistency with the analysis from \cite{Ouchi2010}, we use 2$\sigma$ limits in order to minimise the number of potential interlopers (we show our 3$\sigma$ limits in Table 2). 
Finally, we use the NIR photometry to identify possible low-redshift interlopers which are extremely reddened by dust. For these sources, the Lyman-break is mimicked by the reddening from dust. However, it is possible to identify these interlopers based on their colours red-ward of the narrow-band. We exclude candidates which have  $J-K > 0.5$ (empirically determined as a conservative lower limit) and are detected in the NIR by $>3\sigma$, which results in only 1 additional lower redshift interloper. This source, which is likely an {\sc [OII]} or {\sc [OIII]}/H$\beta$ emitter, has an observed EW of 1050 {\AA}. This additional step is not applied by \cite{Ouchi2010}, but it makes little difference for these luminosities. It is however important for shallower narrow-band surveys, as will be shown in the following sections. The public SpUDS-IRAC catalogue \citep{Kim2011SpUDS} is based on a conservative magnitude limit and therefore does not contain faint enough objects, such that there is no match with any of the LAE candidates within a $2"$ radius, not even \emph{Himiko}. It is however detected in deeper IRAC data \citep{Ouchi2013}. \\
We find a total of 99 Lyman-$\alpha$ candidates in UDS and show the positions in the UDS panel in Fig. $\ref{fig:radec}$. The size of the symbols of the Ly$\alpha$ candidates scales with the logarithm of the line-flux. 

\subsubsection{Lowering the EW criterion}
We retrieve 37 additional LAEs by lowering our excess (EW) criterion to $z-NB921 > 0.5$, but keeping the other conditions fixed. The risk of lowering the EW criterion is that the number of lower redshift interlopers increases. We use the NIR information and photometric redshift to exclude lower-redshift interlopers, because these will generally be very dusty (in order to mimic the Lyman-break) and thus bright and red in the NIR. In \S3.1, we found only 1 interloper if we use an excess criterion of $z-NB921 = 1$. If we lower the criterion to $z-NB921 > 0.5$, we remarkably find only 4 interlopers, where we expected to find more.  However, most of the additional 37 LAEs are very faint in the $z$-band, with magnitudes of $\sim25.5$. This means that they need to have a very red colour ($z-J > 1$ or $z-K > 2 $) to be detected in the near-infrared, since our detection limits are $\sim 24-25$ (see Table 2). Therefore, it is likely that a fraction of interlopers which do not have those extreme colours might be missed. 
Because of their low excess, the additional LAEs have faint luminosities and thus affect mostly the faint end slope in the luminosity function. However, since we do not include these luminosities in the luminosity function due to their low completeness, the specific EW criterion used has little to no effect in our results.

\subsubsection{Comparison to Ouchi et al. (2010)}
The UKIDSS-UDS NB921 data has been analysed by \cite{Ouchi2010}, who found 207 LAE candidates in UDS. The difference of 108 in number of candidates arises since we are even more conservative in our masked regions, limiting magnitude, and visually checking each candidate. However, when applying an analysis similar to Ouchi et al.'s, we find a very similar luminosity function (See \S 5.1 and \S 5.2). From the 16 spectroscopically confirmed LAEs by \cite{Ouchi2010}, we have 15 LAEs in our first selection. The remaining source has a slightly lower excess in our analysis because we do not use {\sc MAG-AUTO}. Even though Ly$\alpha$ is sometimes observed to be more extended than continuum emission \citep[e.g.][]{Steidel2011,Momose2014}, we choose to use aperture photometry for all our measurements. This is motivated by our line-flux completeness estimate in \S 5.1 and a comparison with detection completeness in \S 5.1.1. However, by lowering our excess criterion to 0.9, we also select this source. Since we recover the luminosity function and the spectroscopically confirmed LAEs, we find that our sample of LAEs is in agreement with the sample from \cite{Ouchi2010} and that we fully recover their results.

\begin{table}
\centering
\begin{tabular}{lrr}
\hline    
    {$\bf Field$} & {$\bf Selection \,\,step$} & {$\bf Number\,\, of\, \,sources$} \\ \hline
  \bf UDS & Candidate excess sources & 1514\\
  0.81 deg$^2$ & Spurious/Un-physical & 147 \\
   NB921 $< 26.4$& Line-emitters & 1367 \\ \hline
   & Lyman-break selected & 100 \\
   & Red near-infrared & 1 \\
   & \bf Final LAE candidates  & \bf 99 (15)* \\ \hline
         & & \\   
   \bf COSMOS & Candidate excess sources & 1633\\
  0.96 deg$^2$ & Spurious/Un-physical & 1235 \\
  NB921 $< 25.8$ & Line-emitters& 398 \\ \hline
   & Lyman-break selected & 19 \\ 
   & Red near-infrared & 2 \\
   & Variable sources & 2 \\
   & \bf Final LAE candidates& \bf15 (2)*\\ \hline
         & & \\   
\bf SA22-Deep & Candidate excess sources & 359\\
  0.17 deg$^2$ & Spurious/Un-physical & 2 \\
  NB921 $< 26.4$ &Line-emitters& 357 \\ \hline
   & Lyman-break selected & 6  \\
   & Red near-infrared & 4 \\
   & \bf Final LAE candidates& \bf 2 \\ \hline
      & & \\   

\bf SA22-Wide & Candidate excess sources & 1674\\
  2.72 deg$^2$ & Spurious/Un-physical & 929 \\
 NB921 $< 24.3$  & Line-emitters & 745  \\ \hline
   & Lyman-break selected & 25 \\
   & Red near-infrared & 5 \\
   & Variable sources & 1 \\
   & \bf Final LAE candidates & \bf 19 \\ 
   \hline\end{tabular}
\caption{\small{Number of sources in each selection step, for each field. The final numbers of LAE candidates are printed in bold. *The number between parentheses shows the number of spectroscopically confirmed LAEs. When comparing the fields, it can be seen that there are relatively more line-emitters selected as high-redshift source in UDS. This is because the UDS observations are deeper and the sample therefore exists of fainter observed sources, which generally are at higher redshifts. The number of interlopers identified based on near-infrared colours is relatively high in SA22, where the constraints on the Lyman-break are weaker due to shallower optical photometry. We could only check for variability in the COSMOS field and a small part of the SA22 field.}}
\end{table}

\subsection{Selecting LAEs in COSMOS}
With the photometry presented in \S 2.4, we select line-emitters in the COSMOS field using:\\
\begin{enumerate}
\item  $\bf z-NB921 > 1$\\
\item ${\bf \Sigma > 3}$\\
\item Pass visual inspection\\
\end{enumerate}

With the first two criteria, we find a total of 1633 candidate line-emitters, although immediately exclude 1070 sources for which the excess is unphysically high (see Table 3 for the numbers after each step). We visually inspect the remaining ones and exclude a further 165. The number of candidates excluded based on an unphysical excess or visual checks is considerably higher than in UDS, which is caused by a high number of cosmic rays. The automatic rejection of cosmic rays is less successful because the number of raw exposures is roughly three times lower than in UDS. Other spurious sources have their flux boosted by haloes or spikes caused by bright stars. The difference with UDS is caused by the $z$-band photometry, which is now relatively deeper compared to the narrow-band photometry. Our cut is slightly more conservative in this field than in UDS since we prefer completeness and robustness over the number of sources. We also rely on the UDS sample for the faintest luminosities (because of the deeper imaging) in our determination of the luminosity function and use COSMOS for brighter sources. They agree in number densities (Fig. $\ref{fig:lf_raw}$) indicating that this approach is correct.

In total, we have 398 line-emitters. The photo-z distribution is peaked at {\sc [OIII]}/H$\beta$ at $z=0.83$, with a smaller peak at H$\alpha$ at $z=0.40$ and one around {\sc [OII]} at $z=1.46$. 10 have spectroscopic redshifts, of which 9 are {\sc [OIII]}/H$\beta$  and 1 {\sc [OII]}, see \cite{Sobral2013}. In order to select LAEs, we apply the following criteria to select high redshift line-emitters (see also \S 3.1):
\begin{multline} \nonumber B > 27.9 \land V > 27.3  \land ( i-z > 1.3 \, \, {\vee} \,\, i > 27.0 ) \end{multline}
The optical limits are 2$\sigma$ limits (also applied to UDS) computed by our empty aperture measurements, but consistent with \cite{Muzzin2013}. 
We use the deep near-infrared data to identify dusty lower-redshift interlopers with red near-infrared colours and identify two likely {\sc [OIII]} or {\sc [OII]} emitters, which have $J-K = 1.53$ and $J-K = 1.8$ and observed EW of 420 {\AA} and 350 {\AA}, respectively.  

In addition to this, we are able to check our sources for variability. We have publicly available data from \cite{Sobral2013} which has been taken one year later than the COSMOS NB921 data and reaches a 3$\sigma$ depth of $\sim25$, which is similar to the magnitudes of our faintest Lya candidates. By comparing the magnitudes, we exclude the two brightest candidates (NB921 $\sim 21.5$), because they are completely undetected in the data from \cite{Sobral2013}. This means that their brightness changes by $\sim 5$ magnitudes and that they are thus likely supernovae. This means that similar surveys are expected to have roughly 2 SNe per $0.9$ deg$^2$ as contaminants to their sample of LAEs, except if the data has been split over multiple epochs. This may be very important to interpret results at higher redshift \citep[e.g.][]{Faisst2014,Matthee2014} and shows how important it is to have data spread over time.\\

\begin{figure*}
\centering
\includegraphics[width=16cm]{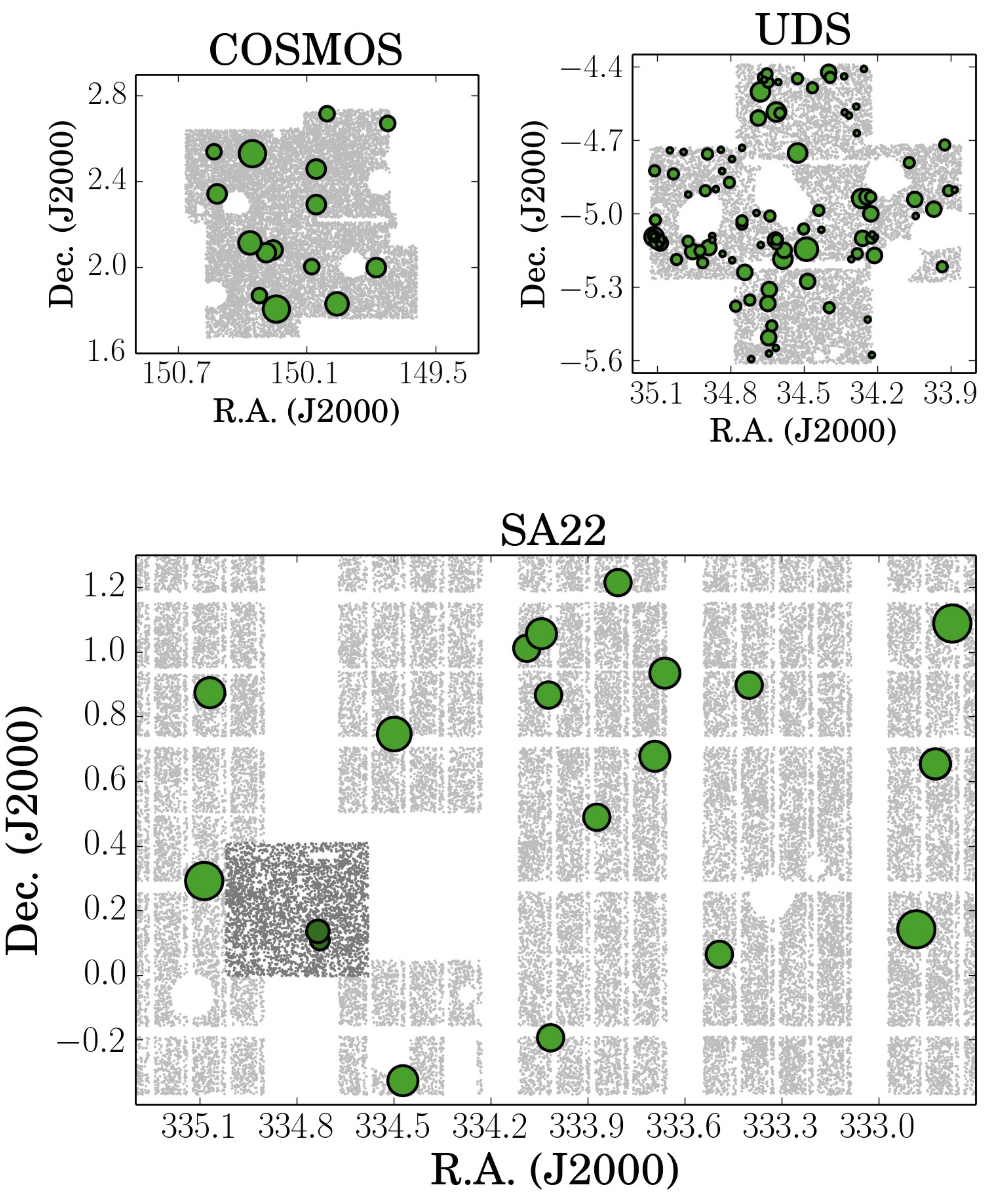}
\caption{\small{Position on the sky of the three survey fields. The three panels show the relative areas of the three fields. Grey dots show all detections with $NB921 < 22$ (chosen in order to control the file size of the images), highlighting the masked regions due to bright stars as empty circles and masked noisy areas which are due to our pointing strategy (such as the grid pattern in SA22). In SA22, the detections in the Deep region have a slightly darker colour. The green symbols show the positions of our LAEs. The size of these symbols scales with the line-flux (luminosity), following ${\rm size} \sim (\rm log_{10}(L_{Ly\alpha}))^{3/5}$, and has the same limits and range for all the three fields. From this, it can be seen that UDS is significantly deeper since it has many more LAEs with small symbols. We furthermore note that two LAE candidates in SA22 are not visible in the image because they overlap with other symbols due to a small separation on the sky ($\sim 4"$).}}
\label{fig:radec}
\end{figure*}

We match our remaining LAE candidates to sources in the S-COSMOS-IRAC catalogue and find one match. This match is our brightest candidate, nicknamed  $``$COSMOS REDSHIFT 7; \emph{CR7}$"$.
After these steps, we find 15 Lyman-$\alpha$ candidates in COSMOS, of which two remarkably bright sources - brighter than \emph{Himiko} -  and show their positions in Fig. $\ref{fig:radec}$. The size of the symbols of the Ly$\alpha$ candidates scales with the logarithmic of the line-flux. \\
We find no additional candidates when lowering the excess criterion. This is because the narrow-band is not as deep as in UDS, while especially the near-infrared constraints are stronger. Therefore, the exclusion of lower redshift interlopers is more successful, especially since we have shown in \S 3.1.1 that the additional candidates are typically very faint, and therefore not present in our slightly shallower COSMOS imaging.

\subsubsection{Spectroscopic follow-up: bright sources in COSMOS}
In UDS, there are 16 spectroscopic confirmed LAEs by \cite{Ouchi2010} using Keck/DEIMOS. For COSMOS, we have obtained spectroscopic follow-up for our brightest two candidates using Keck/DEIMOS and DDT program 294.A-5018 on VLT/X-SHOOTER and VLT/FORS2 presented in \cite{SobralCR7}. Both of these are confirmed Ly$\alpha$ emitters, at redshifts $z=6.604$ and $z=6.541$, respectively.  We nicknamed these galaxies \emph{CR7} (see above) and \emph{MASOSA}\footnote{The nickname \emph{MASOSA} consists of the initials of the first three authors of this paper.}. \emph{CR7} is detected at only half of the narrow-band filter transmission, leading to a $2"$ luminosity of $5.8\times10^{43}$ erg s$^{-1}$, which is a factor 2 higher than our estimate from the photometry. Based on {\sc MAG-AUTO}, the luminosity is  $9.6\times10^{43}$ erg s$^{-1}$, and thus a factor 2.5 brighter than \emph{Himiko}. This is a lower limit since the COSMOS NB921 observations are shallower than in UDS and might therefore miss some lower surface brightness regions and it assumes the $z$-band continuum being flat. It is so extreme that is even detected individually in the $zYJHK$ bands and detected at 5$\sigma$ in the near-infrared stack, with $YJHK = 24.9$. It is also detected in IRAC, with a blue $[3.6]-[4.5]$ colour, consistent with the $[3.6]$ $\mu$m flux being boosted by strong H$\beta$/[OIII] line-emission \citep[e.g.][]{Smit2015}. Because of these broad-band detections, the source can be selected as a Lyman-break galaxy. Indeed, \emph{CR7} is present in the bright $z\sim7$ UltraVISTA catalogs from \cite{Bowler2012,Bowler2014}. However, the Ly$\alpha$ EW is much larger than the values used for the SED fitting (see \cite{SobralCR7}), meaning that the result from their SED fit requires revision and explaining the high $\chi^2$.\\
 \emph{MASOSA} (with a Ly$\alpha$ luminosity of at least $3\times10^{43}$ erg s$^{-1}$, both in a $2"$ aperture and {\sc MAG-AUTO}) is not detected in any of the near-infrared bands, and also not in the stacked image (meaning $YJHK > 26.7$). \emph{MASOSA} is undetected in the $z$-band, meaning that the luminosity is a lower limit. It is only brighter than \emph{Himiko} when measured in $2"$ apertures, but this could also be due to our fainter narrow-band imaging in COSMOS than in UDS. It is not extended (diameter $\sim 0.9"$), while \emph{CR7} and \emph{Himiko} show an extent of $\sim 3"$ in diameter. This means that the sources are of a different nature and therefore interesting targets for follow-up study with e.g. \emph{HST}. \\
None of our LAE candidates are in the zCOSMOS \citep{Lilly2009} or UDSz \citep{Bradshaw2013,McLure2013} catalogues, which we also did not expect, since they would likely be interlopers in that case. As expected, none of our LAE candidates except for \emph{CR7}, are in the UV selected $z\sim7$ galaxy catalogue from \cite{Bowler2014}, since they are not detected in broad-band photometry. None of the other sources in the \cite{Bowler2014} catalogue are selected as line-emitters.

\subsection{Selecting LAEs in SA22 Wide \& Deep}
Using our photometry described in \S 2.4, we select line-emitters using the following criteria:\\
\begin{enumerate}
\item  $\bf z-NB921 > 1$\\
\item ${\bf \Sigma > 3}$\\
\item Pass visual inspection\\ 
\end{enumerate}

After these steps, we find 1674 line-emitters in SA22-Wide, from which we exclude 359 for having an unphysically high excess, and 347 emitters in SA22-Deep (see the corresponding panels in Fig. $\ref{fig:excess}$, and Table 3) and apply the following criteria to select high redshift line-emitters:
\begin{multline} \nonumber u > 26.4 \land g > 26.5  \land  r > 26.2 \land ( i-z > 1.3 \, \, {\vee} \,\, i > 25.9 )  \\
\end{multline}
The optical limits are 2$\sigma$ limits and the $(i-z)$ criterion is similar to that in the other fields.
We visually check each LAE candidate in the NB921 images and exclude a further 560 in SA22-Wide and 2 in SA22-Deep. Most of these objects are cosmic rays which were not detected automatically and sources which have their flux boosted by haloes or spikes caused by bright stars and are thus not real excess sources. In SA22-Wide our $\Sigma$-value from pixel based measurements corresponds to a 3 $\Sigma$ if we measured the background with empty apertures, and to 1.5 $\Sigma$ in SA22-Deep. These differences are caused by a different limiting narrow-band magnitude. \\
There is one LAE candidate in SA22-Wide which is in the small overlapping region with SA22-Deep. This source happens to be variable, as it is undetected in the SA22-Deep data. We could not check the major part of the SA22-Wide field for variability and therefore use a statistical correction to the luminosity function using our empirical results from the COSMOS region, where we found 2 variable sources (likely supernovae) per 0.9 deg$^2$. The area is 2.72 deg$^2$, so we weight our number densities down by 6 sources.\\

Since the optical photometry in SA22 is shallower than in the other fields (see Table 2), there is a higher chance of our candidates being lower redshift interlopers, since the Lyman-break constraints are not as stringent. Using the near-infrared $J$ and $K$ data, we identify objects with significantly red NIR colours ($J-K > 0.5$). In SA22-Deep, we exclude 4 out of our 6 LAE candidates based on optical colours only, and are thus left with 2 LAEs. These four lower redshift interlopers can be called extreme emission line galaxies, since their observed EW is $\sim 400$ {\AA}. In SA22-Wide, we exclude 5 interlopers out of the 24 LAE candidates from the near-infrared photometry. Because the SA22-Wide candidates are brighter, possible interlopers are easier to exclude based on their optical colours. Therefore, this number is relatively lower than in SA22-Deep.
In total, we find 2 LAEs in SA22-Deep, and 19 LAE candidates in SA22-Wide. Their spatial position is shown in Fig. $\ref{fig:radec}$, where the size of the symbols scales with the logarithm of the line-flux.\\

The LAE candidates in SA22-Wide are particularly bright, with luminosities $3-16\times10^{43}$ erg s$^{-1}$, if they are at $z=6.6$. We note that 4 candidates are in pairs which are separated only by $\sim 3-5"$ in the sky (such that only one point is seen in the Fig. $\ref{fig:radec}$). Once these candidates are spectroscopically confirmed, this sample will allow us to study the variety of bright LAEs.

\section{Number counts, completeness and corrections for filter profile bias}
Our main diagnostic is the luminosity function and its evolution with redshift. The luminosity function shows the volume number density of Ly$\alpha$ emitters with a certain (observed) luminosity. In this section, we derive the observed number counts and all other input values for the luminosity function. This is because in addition to the raw number counts per bin, there are important corrections to be made, since our observations and analysis introduce biases and systematic errors, for which we correct in the following subsections. 

The probed volume can be calculated using the FWHM of the filter, since that puts a lower limit and an upper limit to the Ly$\alpha$ emission-line redshift, which are $6.50$ and $6.61$ respectively. We calculate the volume then as the difference in comoving spherical volumes within the upper and lower redshift limits, and multiply this by the fraction of the sky that is our survey area. We find a comoving volume of $9.02\times10^5$ Mpc$^3$ deg$^{-2}$, corresponding to $7.4\times10^5$ Mpc$^3$ in UDS, $8.7\times10^5$ Mpc$^3$ in COSMOS, $1.5\times10^5$ Mpc$^3$ in SA22-Deep and $2.5\times10^6$ Mpc$^3$ in SA22-Wide. In total our volume is $42.6\times10^5$ Mpc$^3$.\\ 

\subsection{Line-flux completeness}
Our selection of line-emitters relies on the measured narrow-band excess and the excess significance. This means that photometric errors can lead to missing real Ly$\alpha$ emitters at $z=6.6$, especially at the faintest luminosities. The result is that our sample is incomplete. How incomplete our survey is at a given luminosity (line-flux) depends on the survey depth, source extraction and selection method.  We can measure the incompleteness with a simulation based on a sample of observed sources which are consistent with being at high redshift ($z>3$, using Lyman-break criteria), but are not detected as line-emitters. This sample is detected and analysed in exactly the same way as our sample of Ly$\alpha$ candidates and has similar narrow-band magnitude distribution. To these sources ($>1,000$ in total in the three deep fields, and $>10,000$ in SA22-Wide) we artificially add line-flux by changing their NB921 and $z$-band magnitude correspondingly, and test whether it is then selected as a line-emitter based on the updated excess and excess significance. The completeness is then obtained for each line-flux by measuring the fraction of sources which is selected as line-emitter after adding the flux \citep[e.g.][]{Sobral2012}. We measure the completeness in the three deep fields for each pointing, but averaged over the SA22-Wide pointings. Because of the limited depth of SA22-Wide, there are less number of sources per pointing and the statistics are thus weaker. We find that the completeness in different pointings in UDS and COSMOS are very similar. We show the line-flux for which each of our fields is complete up to 80 \% in Table 4. These fluxes correspond to luminosities from $6.37\times10^{42}$ erg s$^{-1}$ (UDS) up to $4.9\times10^{43}$ erg s$^{-1}$ (SA22-Wide). Note that even though the narrow-band photometry has similar depths in SA22-Deep and UDS, the line-flux completeness is very different due to a different broad-band limit. This means that a measure of completeness based on detection only will give inconsistent results. \\

For each luminosity bin, we correct the number of sources by dividing by the completeness. We also divide the poissonian errors by this completeness value. Only bins with a completeness higher than 40 \% are included in the luminosity function. The completeness correction is strongest for the faintest luminosities, and thus increases the number density mostly at low luminosities.\\

\begin{table}
\centering
\begin{tabular}{lr}
\hline
{$\bf Field$} & {$\bf 80 \% \,\,completeness \,\,flux$} \\ \hline
UDS & $1.3\times10^{-17}$ erg s$^{-1}$ cm$^{-2}$\\
COSMOS & $4.4\times10^{-17}$ erg s$^{-1}$ cm$^{-2}$\\
SA22-Deep & $7.4\times10^{-17}$ erg s$^{-1}$ cm$^{-2}$\\
SA22-Wide & $10.0\times10^{-17}$ erg s$^{-1}$ cm$^{-2}$\\

\hline\end{tabular}
\caption{\small{The line-flux for which our completeness is 80 \%, shown in our different fields. This depends on both the NB921 and $z$ band depths. Note that, because of this, even though the narrow-band photometry has similar depths in SA22-Deep and UDS, the line-flux completeness is very different due to a different broad-band limit. This highlights the need for line-flux completeness over detection completeness. }}
\end{table}

\begin{figure}
\centering
\includegraphics[width=8.5cm]{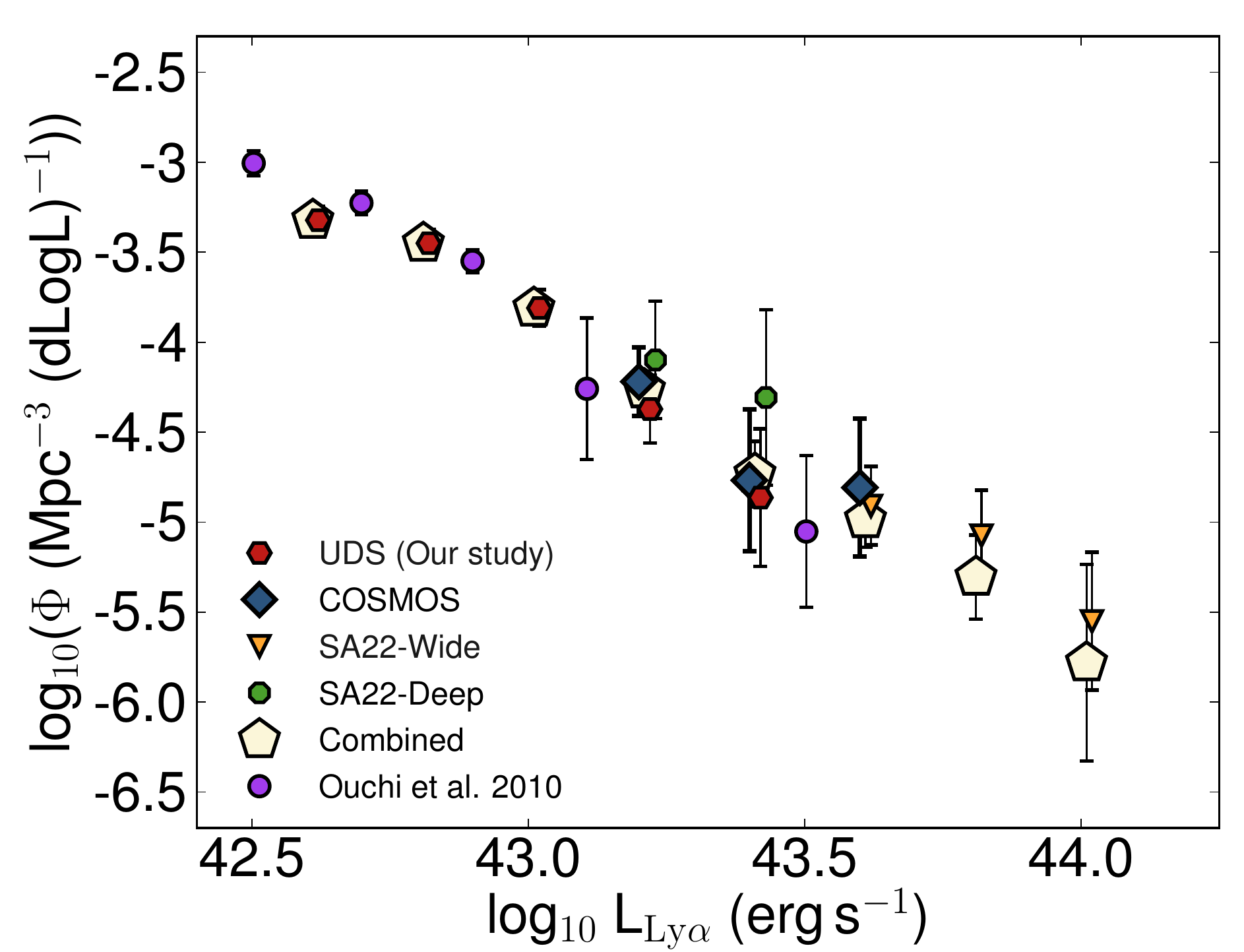}
\caption{\small{Number counts in our three fields, compared to the bins from \citet{Ouchi2010}. The bins are only corrected for completeness. Our bins in UDS vary with this from \citet{Ouchi2010} because we use luminosities based on 2$"$ apertures and apply a different completeness correction. The SA22-Wide bins are corrected for contamination from variable sources and supernovae, empirically calibrated in COSMOS. Note the good agreement between measurements from all fields, although some variance is found which is likely due to cosmic variance.}}
\label{fig:lf_raw}
\end{figure}  

\subsubsection{Detection completeness in SA22}
In order to access the quality of our data in our wide SA22 survey, we use our three spectroscopically confirmed bright LAEs to estimate the detection completeness. By placing them at random positions in our images and see whether we recover them, we know whether our data is sufficient to observe these luminous sources and we can furthermore check our line-flux completeness procedure (see above) and see how the two compare.\\
We produce small cutouts ($5"\times5"$) around \emph{CR7}, \emph{Himiko} and \emph{MASOSA} and add them to 100 random positions per pointing in SA22, excluding masked regions. After this, we run {\sc SExtractor} with identical settings as used on the original images and compute the fraction of our input sources which is detected. We repeat this a 1000 times per image and use the average recovered fraction as detection completeness. On average, we find a detection completeness of 44 \%, with a standard deviation of 20 \% in different pointings. The detection completeness is highest for \emph{MASOSA}, 64 \%, around the average for \emph{CR7}, 43 \%, and lowest for \emph{Himiko}, 27 \%. This is because the first source is not extended, while the other two are extended and therefore have lower surface brightnesses. Note that we do not exclude pixel positions with actual sources or regions with a slightly lower signal to noise, which both decreases the completeness. The average detection completeness is remarkably similar to our estimated line-flux completeness (which is 46 \% for the average line-flux of the three sources). The large variation in detection completeness between the different sources, which have almost the same 2$"$ magnitude, highlights the need for a completeness based on line-flux, instead of detection.\\

\begin{figure}
\centering
\includegraphics[width=8.5cm]{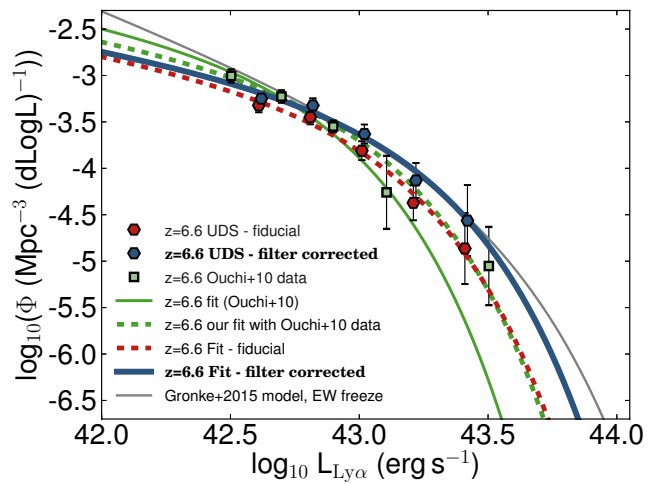}
\caption{\small{Luminosity function at $z=6.6$: comparison of number densities and fits with and without filter profile correction. We compare our number densities in UDS with those of \citet{Ouchi2010}, which are largely based on the UDS field as well. We show that our red hexagons (before correcting for the filter profile) agree well with the green squares from \citet{Ouchi2010}, whose fit to the data is shown as a solid green line. The dashed green line shows our fit to their total data (fixing $\alpha = -1.5$ in Eq. 5), which differs significantly from their published fit. The fit to our data ($\alpha$ fixed to $-1.5$; dashed red line) again agrees well, indicating that our results are similar. The effect of the filter profile correction is shown by comparing the blue hexagons with the red hexagons. The effect is that the number density of bright line-emitters is higher, while the number density of faint line-emitter is slightly lower. The blue line shows the fit to the bins after correcting for the filter profile, which again highlights the effect of the correction. The grey line shows a model prediction by \citet{Gronke2015} which is based on the LBG LF and a Ly$\alpha$ EW distribution, frozen at $z=6.0$. It is remarkable that there it agrees well with the blue curve, despite not being a fit.}}
\label{fig:lf_filter}
\end{figure} 

\begin{table}
\centering
\begin{tabular}{rr}
\hline
{\bf Luminosity bin} & {\bf Number density correction factor} \\ \hline
42.5 & 0.99\\
42.7 & 1.07\\
42.9 & 1.18 \\
43.1 & 1.32\\
43.3 & 1.51\\
43.5 & 1.77\\
43.7 & 2.08\\
43.9 & 2.79\\
\hline\end{tabular}
\caption{\small{Correction factors for the number densities at $z=6.6$. These corrections are made for the bias arising from the observations through the filter profile not being a top-hat. Because of the filter profile, luminous LAEs can be observed as faint LAEs, meaning that their real number densities are higher than observed. This is particularly important for when comparing narrow-band LAE searches with IFU based LAE searches.}}
\end{table}

\subsection{Number densities}
We show our number densities in Fig. $\ref{fig:lf_raw}$ and compare with the number densities from \cite{Ouchi2010} (purple circles), which is based majorly on UDS. Our UDS points agree with those of \cite{Ouchi2010}, while the SA22-Deep and COSMOS bins (which are spectroscopically confirmed) converge at brighter luminosities and are also consistent with \cite{Ouchi2010}. Our SA22-Wide number densities are more uncertain, since there is no spectroscopic confirmation yet and the photometric constraints are weaker than in the other fields. However, even if there are still some contaminants, these further highlight a departure from a Schechter function (already indicated by our spectroscopically confirmed sample) at high luminosities and indicate that the observed Ly$\alpha$ luminosity function at $z=6.6$ can be fitted by a powerlaw (e.g. the pentagons in Fig. $\ref{fig:lf_raw}$). The powerlaw fit is:
\begin{multline*} {\rm log}_{10}(\frac{\Phi}{{\rm Mpc}^{-3}}) = 68.38 - 1.68 \,{\rm log}_{10}(\frac{L_{Ly\alpha}}{\rm erg\, s^{-1}}) \end{multline*} Since we have only two sources in SA22-Deep and since this agrees very well UDS and COSMOS, we will include them when we refer to the UDS+COSMOS sample in the remainder of the text. We will also refer to the SA22-Wide results as SA22. \\

\begin{table*}
\begin{tabular}{lrrrr}
\hline
{$\bf Data-set$} & {$\alpha$ [fixed]} & {$\rm log_{10}(\Phi^*) [Mpc^{-3}]$} & {$\rm log_{10}(L^*) [erg\, s^{-1}]$} & {$\bf \chi^2_{\rm red}$} \\ 

\hline
UDS + COSMOS + SA22 & -2.0 & -4.52$^{+0.10}_{-0.12}$ & 43.61$^{+0.09}_{-0.06}$ & 3.14 \\
  
UDS + COSMOS & -2.0 & -4.13$^{+0.10}_{-0.13}$ & 43.31$^{+0.09}_{-0.65}$ & 2.49\\
UDS & -2.0 & -4.16$^{+0.19}_{-0.44}$  & 43.34$^{+0.38}_{-0.13}$ & 1.78\\
UDS + COSMOS + SA22  without filter correction & -2.0 & -4.40$^{+0.10}_{-0.13}$ & 43.42$^{+0.10}_{-0.07}$ & 1.75 \\

UDS without filter correction & -2.0 & -3.97$^{+0.15}_{-0.21}$ & 43.12$^{+0.15}_{-0.09}$ & 1.31\\ 
 \cite{Ouchi2010} data, our fit & -2.0 & -3.78$^{+0.13}_{-0.18}$  & 43.06$^{+0.13}_{-0.08}$ & 0.48 \\ \hline

 UDS + COSMOS + SA22 & -1.5 & -3.93$^{+0.06}_{-0.05}$  & 43.33$^{+0.04}_{-0.03}$ & 6.65 \\
 UDS + COSMOS & -1.5 & -3.62$^{+0.06}_{-0.06}$ & 43.05$^{+0.06}_{-0.04}$ & 2.07\\
UDS & -1.5 & -3.56$^{+0.10}_{-0.11}$ & 43.01$^{+0.10}_{-0.11}$ & 0.95\\
UDS + COSMOS + SA22 without filter correction & -1.5 & -3.91$^{+0.06}_{-0.07}$  & 43.20$^{+0.05}_{-0.04}$ & 3.87 \\
UDS without filter correction& -1.5 & -3.53$^{+0.09}_{-0.10}$ & 42.88$^{+0.07}_{-0.05}$ &0.68 \\
 \cite{Ouchi2010} data, our fit & -1.5 & -3.35$^{+0.08}_{-0.08}$  & 42.84$^{+0.07}_{-0.05}$ & 0.71\\

\hline\end{tabular}
\caption{\small{Values to our Schechter fits to the luminosity functions. Because of our limited depth, we fix the faint end slope to either $-2$ or $-1.5$ and only keep $\Phi^*$ and L$^*$ as free parameters. }}
\end{table*}

\begin{table*}
\begin{tabular}{lrrr}
\hline
{\bf Data-set} &{\bf $a$} &{\bf $b$} &{\bf $\chi^2_{\rm red}$} \\
\hline
UDS + COSMOS + SA22  & $62.07\pm3.45$  & $-1.53\pm0.08$ & $2.75$\\
UDS + COSMOS & $60.84\pm5.00$ & $-1.50\pm0.12$ & $4.72$ \\
UDS + COSMOS + SA22 before filter correction & $68.38\pm4.58$ & $-1.68\pm0.11$ & $1.62$ \\
UDS + COSMOS before filter correction &$66.26\pm6.10$ & $-1.63\pm0.14$ & $2.67$ \\

\hline\end{tabular}
\caption{\small{Power-law fits to the number densities of the functional form: ${\rm log}_{10}(\frac{\Phi}{{\rm Mpc}^{-3}}) = a + b \,{\rm log}_{10}(\frac{L_{Ly\alpha}}{\rm erg\, s^{-1}})$ The first two are a fit to the corrected number densities (Fig. $\ref{fig:lf_final}$) and the second to the observed (Fig. $\ref{fig:lf_raw}$) number densities. We show both fits with including all three fields, and only for UDS and COSMOS.}}
\end{table*}
Since our LF estimate is based on binning the data, we suffer from Eddington bias \citep[c.f.][]{Ilbert2013}. As luminosities have photometric errors, these uncertainties scatter sources from one bin to the next bin. However, due to the shape of the LF (more sources in the fainter bins than in the bright bins), the luminosity uncertainties move more sources into the luminous end than vice versa. Therefore, this bias tends to overestimate the number of luminous sources and underestimates the fainter sources. It could therefore overestimate the bright end. We are aware of this bias, but do not apply a correction in order for consistency with previous surveys and since its effect are similar between different redshifts.\\

\subsection{Filter profile bias correction}
Since our filter is not a perfect top-hat, the exact redshift of the Ly$\alpha$ emission line influences the observed luminosity. This means that intrinsic luminous LAEs which are detected at the edges of the filter (where the transmission is lower) are observed as fainter LAEs. It also means that the probed volume depends on the luminosity, since luminous sources can be detected over a larger redshift range, but will be observed as fainter sources. For example, our brightest spectroscopically confirmed source in COSMOS, \emph{CR7}, is actually detected at only 50 \% of the transmission and is thus even brighter than our photometric estimate. Corrections for this effect are derived with a simulation, similar to \cite{Sobral2013} for H$\alpha$ line-emitters. We use the Schechter fit of our UDS+COSMOS data to generate a million Ly$\alpha$ emitters and assume that they have a random redshift between the edges of the filter. We then convolve the luminosities with the filter profile into an observed population. Corrections are then obtained by comparing the number of sources in each luminosity bin before and after applying the filter-profile. The result of our correction is that the number-density of luminous sources is increased, while it decreases at low luminosities. We note explicitly that this correction is required to remove the bias from observation strategy, since e.g. an IFU survey without a filter would not suffer from this bias, and that it is not related to any intrinsic effect of the sources. We show the correction factors for $z=6.6$ in Table 5.\\

\begin{figure*}
\centering
\includegraphics[width=16cm]{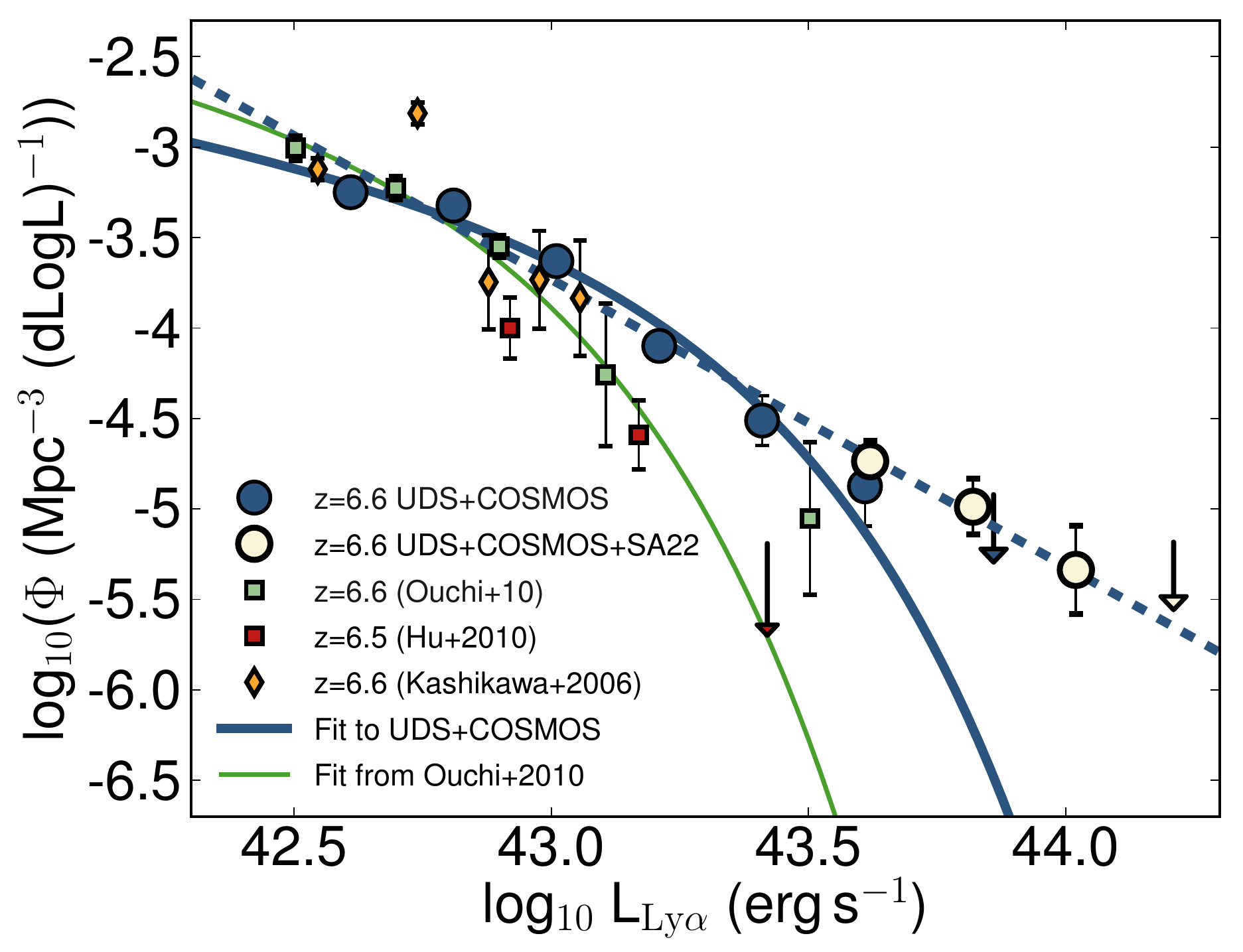}
\caption{\small{The Lyman-$\alpha$ luminosity function at $z=6.6$ compared to literature data. Our most robust luminosity function is shown as a solid blue line. This is a Schechter fit to our combined UDS and COSMOS data (blue circles, see also Table 6), for which the brightest LAEs have all been confirmed spectroscopically. Our additional SA22 data is shown in open circles and is consistent with the upper limits from our robust sample. We also place upper limits (blue and open arrow) at the luminosity bin just brighter than the most luminous observed sources, meaning that there is less than one of these in the probed volume. The dashed blue line is our power-law fit (see Table 7) to the data from all three fields. The fit from \citet{Ouchi2010} at $z=6.6$ differs for two main reasons (see also Fig. $\ref{fig:lf_filter}$), namely practically not including the brightest bin to their fit (due to very large errors, as the fit contains only a single source) and not correcting for different biases caused by the filter profile. This is also the major reason why our results are slightly different with the results from \citet{Kashikawa2006} (ochre diamonds) and \citet{Hu2010} (red squares). Other reasons are cosmic variance, since they only probed small areas (Kashikawa et al.), and small (spectroscopic) completeness (Hu et al.). }}
\label{fig:lf_final}
\end{figure*}  

\subsubsection{Comparison to \citet{Ouchi2010} and effect of the filter profile correction}
There are two reasons for the small differences between our UDS and Ouchi's number densities (Fig. $\ref{fig:lf_raw}$): the first is that our completeness correction is based on line-flux and the selection of emitters, while their completeness correction is based on detection completeness and narrow-band magnitudes. The other difference is that \cite{Ouchi2010} uses {\sc MAG-AUTO} to estimate narrow-band magnitudes (which are used to compute luminosities), while we use the magnitude in 2" apertures. As Ly$\alpha$ is often extended, {\sc MAG-AUTO} might have been a better choice. We however chose to consistently use the 2" aperture since we also used this for our (important) line-flux completeness correction. Using {\sc MAG-AUTO} would mean the introduction of an additional uncertainty. We compare the luminosities derived with both $2"$ apertures and {\sc MAG-AUTO} for the spectroscopic confirmed sample in UDS and find that corrections of +0.11 dex can be used to statistically correct the luminosities. This is used in Fig. $\ref{fig:lf_raw}$ and in all other following luminosity functions. Our results do not strongly depend on this correction. \\

The effect of the filter profile correction on the luminosity function derived by \cite{Ouchi2010} is shown in Fig. $\ref{fig:lf_filter}$. We first fit a Schechter function to our UDS data, both before and after correcting for the profile (respectively the dashed red and solid blue line, see also Table 6) and then compare this fit to the Schechter function from \cite{Ouchi2010}. Because the brightest bin in their sample contains only one source (Himiko), the error on the bin is extreme. Therefore, the fit from Ouchi et al. predicts a 30 times lower number density of bright LAEs than observed. We also fit a Schechter function to their data in log space (green dashed line). Note that we fix $\alpha$ to $-1.5$ in all our fits, similar to previous searches \citep[e.g.][]{Ouchi2008}, since even the deepest UDS data is too shallow to constrain $\alpha$ accurately. However, in Table 6 we also provide fit values for an $\alpha$  fixed to $-2$. From Fig. $\ref{fig:lf_filter}$ it is clear that, first of all, our luminosity function is in agreement with \cite{Ouchi2010} if we use similar corrections and include their brightest bin to the fit (as can be seen by comparing the two dashed lines in Fig. $\ref{fig:lf_filter}$). Second, the effect of the filter profile is highlighted by comparing the solid blue to the dashed red line: after correcting for observational biases from the filter profile, the number density increases mostly at brightest luminosities.

\section{Ly$\alpha$ Luminosity function at $z=6.6$}
\begin{figure*}
\centering
\includegraphics[width=16cm]{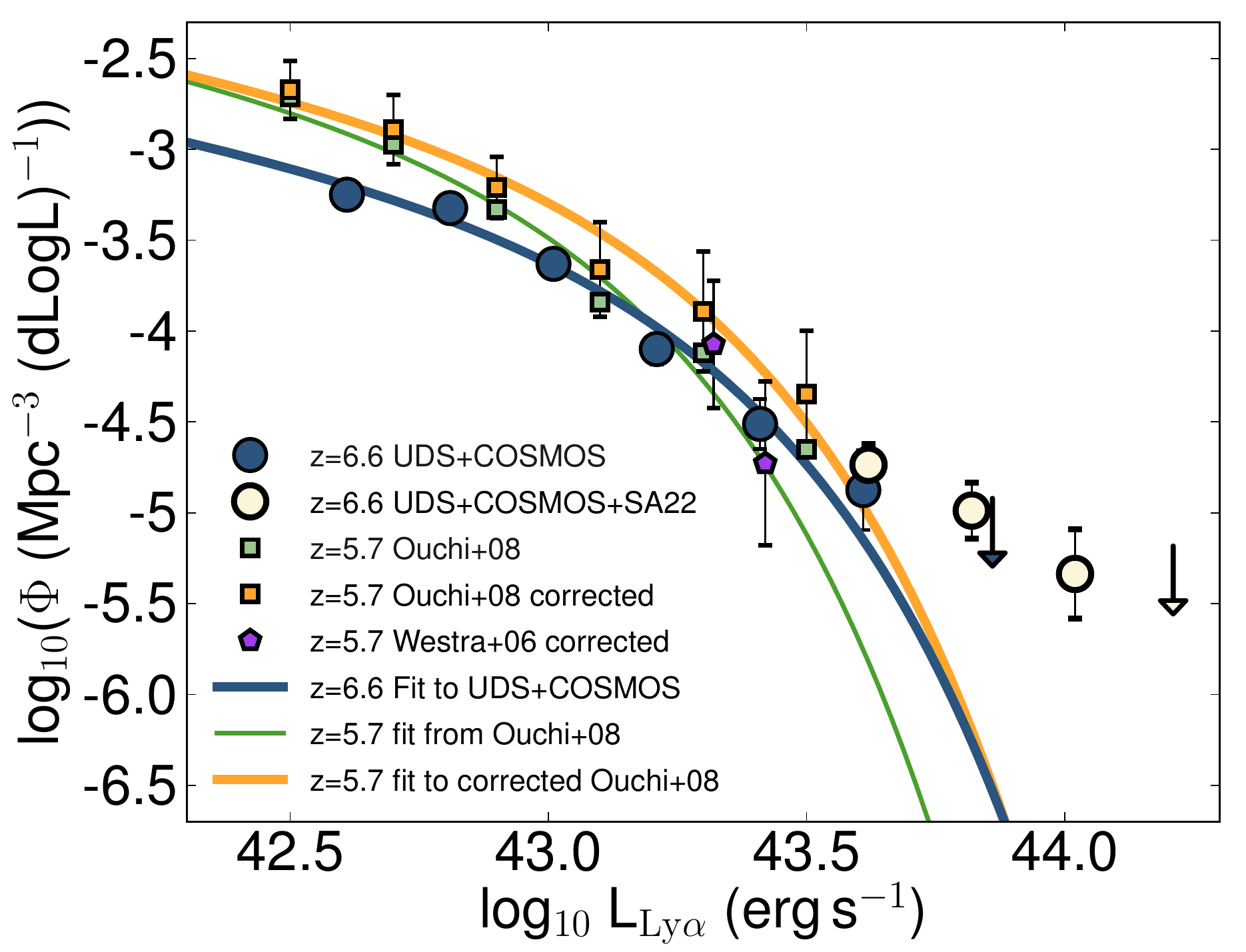}
\caption{\small{Evolution of the Ly$\alpha$ luminosity function from $z= 6.6$ to $z=5.7$. We compare our $z=6.6$ LF (blue solid line) to published data and fits to the data at $z=5.7$. We apply a filter profile correction to the data from \citet{Ouchi2008} at $z=5.7$ (green: uncorrected, ochre: corrected) and fit a Schechter function to the corrected data (orange solid line). We also show the spectroscopically confirmed results from \citet{Westra2006} at $z=5.7$. Comparing the orange solid line to the blue solid line shows that the LF evolves only at the faint end and not (as has been claimed by \citet{Ouchi2010}) at all luminosities. In fact, we find no evolution for $L > 10^{43.5}$ erg s$^{-1}$. We show in \S 6.2 that this may be a consequence of re-ionisation. We do not show our fit which includes the SA22 data-points, since there is no comparison available at $z=5.7$. This motivates the need for larger volumes at $z=5.7$ as well, in order to see if there is evolution at the bright end. }}
\label{fig:lf_evo}
\end{figure*}

In this section, we present the $z=6.6$ Ly$\alpha$ LF from our combined analysis in UDS, COSMOS and SA22.  As a functional form, we use the well-known Schechter function:
\begin{equation}
\phi(L)dL = \phi^* (L/L^*)^{\alpha}\,{\rm exp}(-L/L^*)d(L/L^*)
\end{equation}	
We convert our observed line-fluxes to luminosities by assuming a luminosity distance corresponding to a redshift of 6.56, which is the redshift of the center of the filter. We combine the luminosities in bins with widths of 0.2 dex and count the number of sources within each bin and correct this number for incompleteness. The errors on the bins are taken to be Poissonian. The number of sources is divided by the probed volume, such that we obtain a number density. We then apply our corrections for the filter profile bias. Only data where the completeness is at least 40 \% is included. The resulting luminosity function is shown in Fig. $\ref{fig:lf_final}$, where we also compare with other published $z=6.6$ LAE data. The evolution between $z=5.7$ and $z=6.6$ is shown in Fig. $\ref{fig:lf_evo}$, while the left panel of Fig. $\ref{fig:lf_reion}$ shows the evolution towards $z=7.3$. We are cautious about interpreting the results from SA22 because of the less stringent photometric criteria, even though they fully agree with results from the other fields. The results from UDS and COSMOS however, are confirmed by spectroscopy. \\

It is interesting to compare these results with the model from \cite{Gronke2015}, which is shown as the black line in Fig. $\ref{fig:lf_filter}$. This model uses the UV LF and a probability distribution (PDF) of Ly$\alpha$ EWs to predict the Ly$\alpha$ LF. The EW distribution generally evolves with redshift, but in this case, it is frozen to the EW PDF at $z=6.0$, because of possible effects from re-ionisation. It is remarkable that the prediction from \cite{Gronke2015} seems to be consistent with our blue points. Differences arise because of their steeper faint end slope ($\sim -2.2$), which is largely unconstrained by the depth of our current data of LAEs. The agreement highlights the need for the correction of the filter profile bias when comparing narrow-band derived LFs with LFs derived from spectroscopy (either follow-up or blind IFU).\\

As noted in \S 5.2.1, our results in UDS differ by those from \cite{Ouchi2010} at brighter luminosities due to a different treatment of the brightest bin in the fit (solid green line) and by correcting for the filter profile (which effect is shown in Fig. $\ref{fig:lf_filter}$). This explains also the differences (although largely within the errors) with \cite{Kashikawa2006}, although cosmic variance plays a role because of their limited survey area. As noted by \cite{Kashikawa2011}, the difference between the results from \cite{Hu2010} and the others is due to incompleteness of the sample of \cite{Hu2010}, since they rely on spectroscopic follow-up with too short integration times.
Before correcting for the filter profile, our results in COSMOS agree with those from \cite{Ouchi2010} (see Fig. $\ref{fig:lf_raw}$), but even after correcting for the filter profile, our combined UDS+COSMOS LF agrees with the brightest bin of \cite{Ouchi2010}, and disagrees only slightly with the second brightest bin. Our SA22 results are not yet confirmed spectroscopically and are thus upper limits when viewed most conservatively. There is however excellent agreement with the spectroscopically confirmed COSMOS sources and with the upper limits from the UDS+COSMOS sample (blue arrow in Fig. $\ref{fig:lf_final}$). If all (or even only a fraction) of these very bright Ly$\alpha$ emitters in SA22 are confirmed, this indicates that the observed Ly$\alpha$ luminosity function at $z=6.6$ can be fitted by a powerlaw (e.g. the pentagons in Fig. $\ref{fig:lf_raw}$), similar to the UV luminosity function at $z=6-7$ \citep[e.g.][]{Bowler2014}. \\

Now we have used a combination of wide and ultra-deep fields, we have established a new LF at $z=6.6$ (see Table 6) and will compare this with results in the literature at lower redshift. To be conservative, we will use our fit to the UDS+COSMOS sample for comparison, although our results are only strengthened by the results from SA22. 

\subsection{The $\bf z=5.7-6.6$ evolution of the LF}
Previous studies \citep[e.g.][]{vBreuk2005,Ajiki2006,Shimasaku2006,Gronwall2007,Ouchi2008} have found that the observed Ly$\alpha$ luminosity function is remarkably constant between $z=3-6$. While the LBG LF declines over this redshift range, the implication is that the strength of the Ly$\alpha$ line increases with redshift, which is also confirmed spectroscopically \citep[e.g.][]{Stark2010}. At $z>6$, there is evidence both from a declining success-rate of spectroscopic observations of LBGs \citep[e.g.][]{Schenker2012}, and also for a drop in the Ly$\alpha$ LF \citep{Ouchi2010,Konno2014}. In our independent analysis we confirm evolution of the Ly$\alpha$ LF from $z=5.7$ to $z=6.6$, although only robustly at fainter luminosities, $L<10^{43}$ erg s$^{-1}$, see Fig. $\ref{fig:lf_evo}$.
As a comparison at $z=5.7$, we use data-points from \cite{Ouchi2008} (green squares in Fig. $\ref{fig:lf_evo}$), complemented with the wide area survey from \cite{Westra2006}. We were not able to compare our results to the number densities of \cite{Murayama2007} at $z=5.7$ (2 deg$^2$ in COSMOS) since they do not correct for completeness. As discussed above, we apply a correction for the filter profile, based on a generated sample of a million LAEs following the $z=5.7$ Schechter function from \cite{Ouchi2008}. The result is shown as the orange squares in Fig. $\ref{fig:lf_evo}$. For the data from \cite{Westra2006}, we only show the corrected bins. We then fit a new Schechter function to the corrected points, shown as a solid orange line. The evolution between $z=5.7$ and $z=6.6$ can be seen by comparing the orange and blue solid lines. It can be seen that at fainter luminosities (L$_{Ly\alpha} \sim 10^{42.5}$ erg s$^{-1}$) the observed number density of LAEs declines by a factor $\sim 3$, while there is no evolution at the bright end L$_{Ly\alpha} > 10^{43.5}$ erg s$^{-1}$). Note that we only compare to our $z=6.6$ sample based on UDS and COSMOS and that the results at the bright end are confirmed spectroscopically. Addition of the SA22 bins only strengthens the conclusion. Unfortunately, however, it is not yet possible to compare the SA22 results with lower redshifts, since there is no comparable data-set at $z<6$. Even without the filter profile correction, we find differential evolution of the luminosity function. We will discuss explanations for this differential evolution in the discussion in \S 6.1 and \S 6.2. 
When adding even higher redshift data from $z=7.0$ \citep{Iye2006,Ota2010} and $z=7.3$ \citep{Shibuya2012,Konno2014} narrow-band surveys (see the left panel of Fig. $\ref{fig:lf_reion}$), the LF keeps decreasing at luminosities smaller than $10^{43}$ erg s$^{-1}$. Note that these bins consist of just a handful of sources and that we have not corrected them for the bias due to the filter profile. As the survey areas of these surveys are below 0.3 deg$^2$, they are all limited severely by cosmic variance and it is therefore not possible to compare the evolution at the bright end. A few additional shallower pointings will be useful to constrain the evolution of the bright end as well, and confirm or neglect a continuing differential evolution.

\section{Discussion}
\subsection{LF evolution and re-ionisation}
The evolution of the observed Ly$\alpha$ LF can be caused by different processes: \\
i) an intrinsic dimming of the Ly$\alpha$ EW. As the cosmic star formation rate density derived from UV observations at $z>6$ continues to decline with redshift \citep[e.g.][]{Bouwens2014,McLeod2014}, the production rate of Ly$\alpha$ can decrease due to a lower star formation rate, since Ly$\alpha$ is emitted by the recombination of hydrogen atoms which are photo-ionised by massive, young, short-lived stars. However, the cosmic star formation rate is already declining from at least $z>3$, while the observed Ly$\alpha$ LF is constant between $z=3-6$, indicating an increase in Ly$\alpha$ EW with increasing redshift (see also \cite{Stark2010}) and it is hard to explain a sudden reversal of this trend. Furthermore, the star formation towards higher redshifts can be partly contributed by formation of metal free Pop III stars, which are predicted to produce copious amounts of Ly$\alpha$ emission \citep{Schaerer2003}. 

ii) the observed drop in the Ly$\alpha$ LF can be explained by a lower escape of Ly$\alpha$ in the interstellar medium of galaxies \citep{Dayal2012,Dijkstra2014}. This escape fraction is largely unconstrained, although first direct measurements of matched H$\alpha$-Ly$\alpha$ observations at $z=2.2$ indicate an average escape of 5 \% \citep{Hayes2010}. However, other, more indirect measurements of the escape fraction in LAEs is in general higher \cite[30 \%, e.g.][]{Wardlow2014,Kusakabe2014}. Using a joint analysis of the evolution of the Ly$\alpha$ and LBG luminosity functions, \cite{Hayes2011} finds that the volumetric (statistical) Ly$\alpha$ escape fraction increases with redshift up to $z=6$ and decreases at higher redshift, although this result is based on the integration of the Ly$\alpha$ LF from \cite{Ouchi2010}. As we show, in this paper, their large error on the bright end results in an underestimate of the luminosity density. The intrinsic Ly$\alpha$ escape fraction can be measured directly at $z>3$ once infrared spectroscopy of the H$\alpha$ line with the \emph{James Webb Space Telescope} is possible. It is thought that the primary factor in driving the escape fraction is the neutral hydrogen column density \citep[e.g.][]{Nakajima2012}, but kinematics and dust production can also have a role. Similar to the intrinsic dimming of Ly$\alpha$ EW, it is hard to explain a reversal of the trend highlighted by \cite{Hayes2011}, i.e. an increase up to $z\sim 6$ and a decline afterwards. 

\begin{figure*}
\begin{tabular}{cc}

\includegraphics[width=8.5cm]{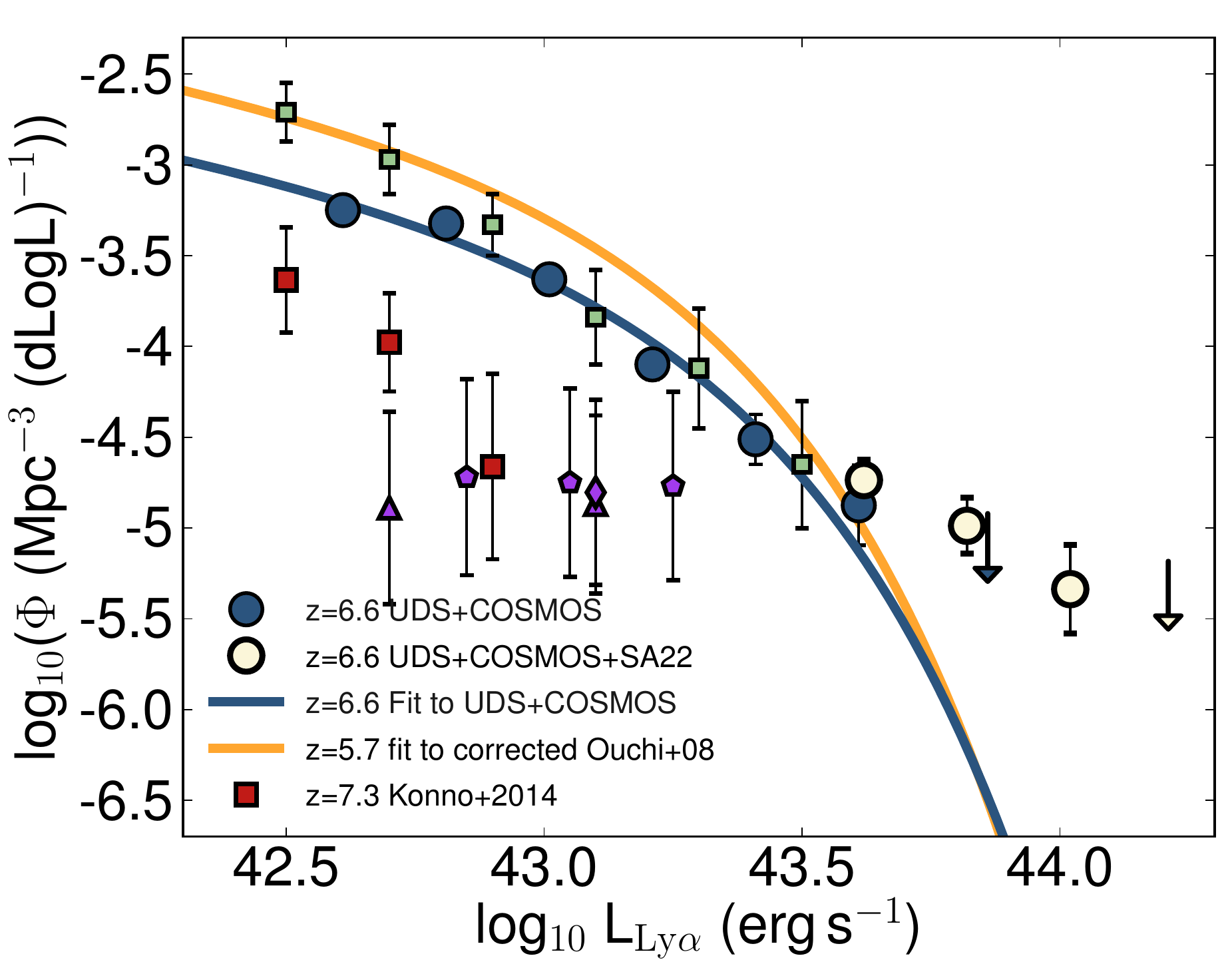}&

\includegraphics[width=8.5cm]{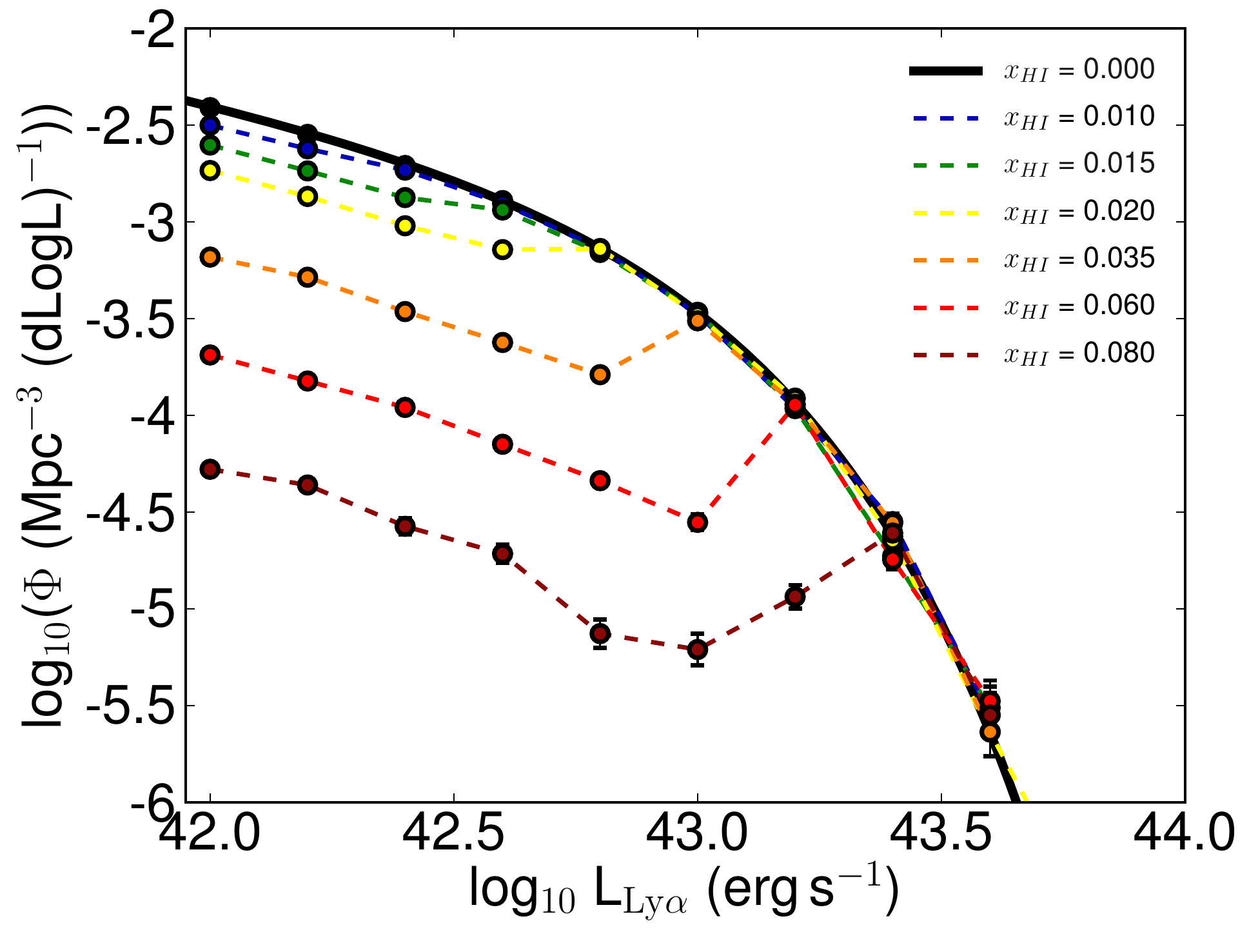}\\

\end{tabular}
\caption{\small{{\textit {Left:}} Evolution of the luminosity function evolution from $z= 7.3$ to $z=5.7$. We compare our $z=6.6$ LF (blue solid line) to published data at $z>7$ \citet{Konno2014} (red squares) \citet{Shibuya2012,Ota2010,Iye2006} (purple triangles, pentagons and diamond, respectively). We also show our LF fit to the corrected $z=5.7$ data (orange solid line) and the green squares show the number densities at $z=5.7$ from \citet{Ouchi2008}. Note that the errors on the $z>7$ data are still significant and that the surveys are limited to small areas, meaning that the LF is still unconstrained at luminosities $>10^{43.3}$ erg/s. A few additional, shallower pointings with the Subaru S-Cam and the NB101 ($z=7.3$) filter would place useful constraints on the evolution of the Ly$\alpha$ LF at these epochs. {\textit {Right:}} Toy-model evolution of the luminosity function in a neutral IGM. The black line shows the input LF, which is the $z=5.7$ LF from \citet{Ouchi2008}. We fix the parameters controlling the age and number of escaping ionising photons from LAEs (to age = 100 Myr, $f_{esc, ion}$ = 5\% and $f_{esc, Ly\alpha}$ = 30 \%) and investigate how a changing neutral fraction ($X_{HI}$) influences the observed LF. As can be seen, the evolution starts at the faintest luminosities only and gradually towards higher luminosities. The highest luminosities ($>10^{43.5}$ erg s$^{-1}$) are still observable. Faint LAEs are observable if they are in the ionised sphere of a neighbouring, luminous source. Therefore, the observed faint-end slope depends on the clustering. The turnover luminosity, at which LAEs can ionise their own surrounding enough, depends on our input parameters and the neutral fraction. Note that our specific values for the neutral fraction are arbitrary since our input parameters are largely undetermined, and they are only showed for illustration purposes. While typical models require much larger changes in the neutral fraction (e.g. \citet{DijkstraReview}), our values are particularly low because we assume that all Ly$\alpha$ emission is absorbed when the ionised sphere is $<1$ Mpc, while in reality only part of the line-emission is absorbed, and we ignore other processes such as galaxy outflows which can increase the observability of Ly$\alpha$. }}
\label{fig:lf_reion}
\end{figure*}

iii) a sudden decrease in the observability of Ly$\alpha$ can possibly be explained by an increase in the incidence of dense pockets of neutral hydrogen in the line of sight (e.g. Lyman Limit Systems or Damped Lyman-$\alpha$ absorbers). \cite{BoltonHaehnelt2013} argue that the majority of optically thick gas along lines of sight is present in absorption systems, although again, there is no definite explanation why there should be a sudden increase, or why this increase depends strongly on Ly$\alpha$ luminosity. 

iv) the most quoted reason \citep[e.g.][]{Santos2004,HaimanCen2005,McQuinn2007,Choudhoury2014,Jensen2014} is that the observed decrease in both the evolution in the Ly$\alpha$ LF and the declining spectroscopic success-rate in LBG follow-up, is caused by an increased neutral fraction of the IGM, which leads to a higher opacity to Ly$\alpha$ photons. \cite{Mesinger2015} however argue that re-ionisation can not be the only driver of the observed changing opacity to Ly$\alpha$ and \cite{Konno2014} mention that the neutral IGM fraction required to explain their observed evolution of the Ly$\alpha$ LF (see left panel of Fig. $\ref{fig:lf_reion}$) is in tension with results from the polarisation of the Cosmic Microwave Background (CMB) \citep{Planck2014}, with a mean re-ionisation redshift of $z=11.1$. On the other hand, \cite{Choudhoury2014} argue that the most recent CMB results are not in disfavour with a re-ionisation epoch that ends between $z=6-7$. The latest Planck results suggest a re-ionisation redshift of $z=8.8$ \citep{Planck2015}, but this result is model dependent.  

A lot of these explanations are very degenerate with the currently available observational data and can only be solved by disentangling the effect from galaxy formation (e.g. a varying intrinsic production or escape of Ly$\alpha$ photons) with a cosmological cause (re-ionisation of the IGM). One way to overcome this is to study the clustering properties of LAEs in wide area surveys, as is the goal of the HyperSuprimeCam survey on the Subaru telescope (see also \S 6.2). \cite{Jensen2014} show that an increased (observed) clustering signal of LAEs is indicative of a more neutral IGM, because Ly$\alpha$ is preferentially observed in an ionised region and more clustered LAEs will lead to a larger ionised region around them. Mimicking this clustering signal is unlikely by changes in the intrinsic escape fraction or production of Ly$\alpha$. Another measurement which will overcome the degeneracy is to measure the intrinsic Ly$\alpha$ escape fraction directly using spectroscopic H$\alpha$ measurements, for example possible with a matched narrow-band survey to Ly$\alpha$ with a wide field camera which is sensitive to 2-5 micron radiation in space (since the night sky background is very high in these wavelength regions). 

In the next section, we argue that our observed differential evolution of the Ly$\alpha$ LF can be explained by a simple model for re-ionisation, although we do not claim that this is the only possibility. 

\subsection{Toy-model for the evolution of the Ly$\alpha$ LF during re-ionisation}
Since the process of re-ionisation likely has a patchy nature, where the Universe is ionised either bottom-up (overdense, early collapsing regions first) or top-down (voids first), there is an observable effect in the evolution of the Ly$\alpha$ LF. \cite{HaimanCen2005} argue that the observability of bright LAEs is less attenuated by a neutral IGM than faint LAEs, since the ionised spheres around these are typically larger. This indicates that the evolution of the LF happens mostly at the faint end, where a smaller neutral fraction is sufficient to prevent the observability. This scenario is similar to our observations: most of the evolution between $z=5.7$ and $z=6.6$ happens at fainter luminosities, see Fig. $\ref{fig:lf_evo}$. 

We use the following toy-model to show that a differential evolution indeed follows from some basic assumptions about how the process of re-ionisation happened. We first assume that the intrinsic Ly$\alpha$ LF at $z=6.6$ (without IGM absorption) is similar to the $z=5.7$ LF from \cite{Ouchi2008}. Using this LF, we generate a sample of LAEs with a minimum luminosity $10^{41.5}$ erg s$^{-1}$ and consider three cases: First, we place them at random positions in our survey area, but also vary this to clustered positions, where the clustering has a functional form varying from $`$low' clustering (powerlaw) to $`$high' clustering (exponential). In case of the powerlaw clustering, the probability $p(x)$ that a LAE has a neighbouring LAE within a distance $x$, scales with $p(x) \sim - x$, while this scales with $p(x) \sim e^{-x}$ in the case of exponential clustering. This means that for the powerlaw clustering, the expected number of neighbours within 1 Mpc is 2 times higher than based on a random spatial distribution, while it is 13 times higher for the exponential clustering. 

Assuming that the LAEs are the only source of an ionised region around them (without any contribution from fainter neighbouring sources), we use the following formula for the radius of their Str\"omgren sphere at $z=6.56$ from \cite{CenHaiman2000}:
\begin{equation}
R_S = 4.3 \,x_{\rm IGM}^{-1/3} \,(\frac{f_{esc, ion} N_{\gamma, em}}{1.3\times10^{57} \rm{s^{-1}}} )^{1/3} \,(\frac{t}{2\times 10^7 \rm{yr}})^{1/3} \,\,\rm{Mpc}
\end{equation}

\noindent where $x_{\rm IGM}$ is the neutral hydrogen fraction in the IGM, $f_{esc, ion}$ the escape fraction of ionising radiation, $N_{\gamma, em}$ the number of ionising photons per second and $t$ the time the emitter has been emitting ionising radiation. Under basic assumptions, we calculate the number of ionising photons by converting the Ly$\alpha$ to H$\alpha$ luminosity with:

\begin{equation}
L(H\alpha)=8.6 f_{esc, Ly\alpha} L(Ly\alpha) \,\,\,{\rm erg}\,\, {\rm s}^{-1}
\end{equation}

\noindent Here, $f_{esc, Lya}$ is the intrinsic Ly$\alpha$ escape fraction and the 8.6 value corresponds to case B recombination \citep{Osterbrock1989}. Under the same case B recombination assumption and using a Salpeter IMF, the H$\alpha$ luminosity is then converted to the number of ionising photons using \citep[e.g.][]{Orsi2014}:

\begin{equation}
N_{\gamma, em} = \frac{L(H\alpha)}{1.37\times10^{-12}}  \,\,\,{\rm s}^{-1}
\end{equation}
 
According to \cite{CenHaiman2000}, it takes an ionised sphere of roughly 1 Mpc for Ly$\alpha$ to redshift out of resonance wavelength. Therefore, we use the above formalism to compute the ionised regions around LAEs, which can overlap if the sources are clustered, and check which LAEs can be observed, by being able to travel at least 1 Mpc through an ionised region through any sightline. As an output, we compile the observed Ly$\alpha$ LF and compare this to the input. Our results depend obviously on our choice for the Ly$\alpha$ escape fraction, the age of LAEs and the neutral fraction. However, we are only interested in qualitatively investigating the effect of this toy-model on the observed evolution of the LF, so the specific values are not that important, also because they all have the same powerlaw-scaling in Eq. 6. Our simulation likely breaks down when there is a strong dependence of intrinsic Ly$\alpha$ luminosity on age, escape fraction or clustering (halo mass), since this would affect the relative sizes of ionised spheres and thus the observability of LAEs. \\

After running our simulation 2000 times (to overcome sampling errors) for each set of clustering strengths and other input parameters, we find that the effect is that without clustering, brighter LAEs are more likely to be observed, since they are powerful enough to ionise their own surroundings. Changing any of the parameters except for clustering only varies the intrinsic Ly$\alpha$ luminosity at which a LAE is able to ionise its surrounding enough. A stronger clustering leads to an increase in the number of observed faint LAEs, since they likely reside in ionised regions of larger LAEs, or are able to ionise a large enough region with some close neighbours themselves.\\ 

Our results of the simulation are illustrated in the right panel of Fig. $\ref{fig:lf_reion}$. In this Figure, we fixed the parameters $t$, $f_{esc, ion}$ and $f_{esc, Ly\alpha}$  to 100 Myr, 5 \% and 30 \%, respectively, and also fixed the clustering (to the maximum, exponential clustering strength) and only vary the neutral fraction. Our input LF (which is the same in the fully ionised case) is the LF at $z=5.7$ from \cite{Ouchi2008}. One can see that once the neutral fraction increases, the majority of the evolution happens at luminosities below $10^{43}$ erg s$^{-1}$, although the exact position depends on the specific neutral medium. This turnover point also depends on our choice of input parameters and therefore, since these are all uncertain, we do not use our model for estimates of the neutral fraction at $z=6.6$. The turnover point corresponds to the luminosity at which a LAE is able to create a Str\"omgren sphere of 1 Mpc. However, even in the most neutral fraction, there are still faint LAEs which can be observed. This is due to clustering, since they are located in the ionised regions of larger LAEs. The faint-end slope depends on the clustering, being the strongest if ionising sources are highly clustered. Concluding, the right panel of Fig. $\ref{fig:lf_reion}$ shows that in our simple toy-model for how re-ionisation happens and affects the observability of LAEs, the differential evolution that we observe between $z=5.7$ and $z=6.6$ can be explained by a higher neutral fraction of the IGM. It also shows that towards higher redshifts, where the Universe keeps becoming more neutral, strategies aiming at detecting LAEs benefit more from a wide-field approach, since LAEs are easier to be observed as they are able to ionise their own surroundings sufficiently.\\

Note that our specific values for the neutral fraction are arbitrary since our input parameters are largely undetermined. We therefore do not aim to use our model to quantitatively measure the neutral fraction, but to qualitatively explain our observations. While typical models require much larger changes in the neutral fraction to significantly reduce Ly$\alpha$ emission \citep[e.g.][]{DijkstraReview}, our values are particularly low. This is because we assume that all Ly$\alpha$ emission is absorbed when the ionised sphere is $<1$ Mpc, while in reality only the blue part of the line-emission is absorbed, and the red wing broadens due to resonant scattering. We also ignore other processes which can increase the observability of Ly$\alpha$, such as galaxy outflows.

The observed differential evolution for Ly$\alpha$ selected galaxies is also seen in the spectroscopic follow-up of Lyman-break galaxies. For example, \cite{Ono2012} finds that the drop in the fraction of UV selected galaxies which are detected in Ly$\alpha$ at $z>6$ is stronger for UV faint galaxies, using a compilation of surveys. This is particularly interesting as at lower redshift, UV faint galaxies tend to have a higher Ly$\alpha$ fraction \citep[e.g.][]{Stark2010}. As mentioned by \cite{Ono2012}, this differential drop indicates that re-ionisation happens first in over-dense regions and is qualitatively consistent with our observations of the Ly$\alpha$ LF.

\subsection{Future surveys}
We can use our number counts, shown in Fig. $\ref{fig:lf_raw}$, to estimate the expected number of sources in future surveys such as the extragalactic Hyper Suprime-Cam (HSC) survey, which the Subaru telescope will undertake. Due to its large field of view (1.5 deg$^2$), the HSC is suited excellently for pushing to a survey with a similar depth as the narrow-band data in UDS and a $>10$ deg$^2$ area, and even wider areas for slightly shallower surveys. From our spectroscopic confirmed sample only (and without the filter correction), we find that the observed number density at $5\times10^{43}$erg s$^{-1}$, the luminosity of our brightest source in COSMOS, is roughly 30 times higher than based on the LF from \cite{Ouchi2010} (see Fig. $\ref{fig:lf_filter}$). The HSC survey\footnote{www.naoj.org/Projects/HSC/surveyplan.html} has a planned deep component of $\sim 30$ deg$^2$ (part of which is SA22) to a NB921 depth of 25.6 and an ultra-deep component (in UDS and COSMOS) of $\sim 3.6$ deg$^2$ to a depth of 26.2. If we use our powerlaw fit from \S 5.1, we therefore expect that $\sim 60$ bright LAEs ($>5\times10^{43}$erg s$^{-1}$) will be found in total. These bright sources will be excellent targets for follow-up with \emph{JWST}, to study the Ly$\alpha$ escape fraction directly with H$\alpha$ measurements, and to study the metallicity and ionisation state with other nebular emission lines. The HSC survey will also be able to study our brightest LAEs to more detail, because it will be able to observe the more extended, lower surface brightness regions because of deeper survey limits. Furthermore, the HSC survey will also obtain additional wide observations with the $z=5.7$ Ly$\alpha$ filter, such that the SA22 results at the highest luminosities can be compared. 

Another survey which recently started observations is the \emph{Javalambre Physics of the Accelerating Universe Astronomical Survey (J-PAS)} \citep{Benitez2014}, which will survey 8500 deg$^2$ with 54 narrow-band filters, from 3500 {\AA} up to a wavelength of 10000 {\AA}, to a depth of $\sim $ 22.5. Its goal is to study dark energy at redshifts $z<1$, but it is also interesting for extragalactic studies of quasars and LAEs due to its very wide area and large set of narrow-band filters. If we use our powerlaw fit (Table 7) and extrapolate it to a luminosity of $10^{45}$ erg s$^{-1}$ (e.g. the Ly$\alpha$ luminosity of the $z=7.085$ quasar \citep{Mortlock2011}), we estimate that it will find 230 bright LAEs per NB filter. A caveat however is that excluding interlopers is challenging with the depth of the optical imaging and without the availability of near-infrared.

The brightest LAEs detected with optical surveys (e.g. up to $z \sim 7.3$) are the best comparison with samples of even higher redshift LAEs. To date however, no LAE has been detected in the near-infrared \citep[e.g. $z=7.7$ and $z=8.8$;][]{Faisst2014,Matthee2014}, but upcoming surveys will increase both the sensitivities (which is a main difficulty due to the high sky background) and probed volumes. Nevertheless, our results, showing little to no evolution at the bright end, encourage further searches for the brightest LAEs, at least at $z\sim7-8$. Eventually, the Euclid satellite will perform a deep spectroscopic survey over a wide area (40 deg$^2$, $5\times10^{-17}$erg cm$^{-2}$  s$^{-1}$) in the near-infrared, which will detect bright $z>7.3$ LAEs if they exist. 

\section{Conclusions}
The main conclusions of this work are:

\begin{itemize}
\item Using a combination of wide and ultra-deep surveys in SA22, UDS and COSMOS with the NB921 filter on the Subaru telescope, we obtain a large sample of LAEs at $z=6.6$, spanning the largest dynamical range of luminosities to date, and derive a new luminosity function, which overcomes cosmic variance at the bright end because of our large and independent volumes.
\item We identify lower redshift interlopers (including extreme emission line galaxies at $z=1-2$) with near-infrared data. This way we find a number of extreme emission line galaxies with rest-frame EWs of $>400$ {\AA}.
\item In total, we find 99+15+19  (UDS, COSMOS, SA22) LAE candidates at $z=6.6$. Of these, 18 are confirmed spectroscopically. 16 in UDS by \cite{Ouchi2010} and 2 in COSMOS by \cite{SobralCR7}. These two sources are the brightest LAEs at $z=6.6$ so far and confirm that sources with a luminosity similar to \emph{Himiko} are not as rare as previously thought, and have number densities of $\sim 1.5^{+1.2}_{-0.9}\times10^{-5}$ Mpc$^{-3}$.
\item In our wide-field shallow SA22 data-set, we find 19 LAEs which are even brighter than the ones in UDS and COSMOS, although these have no spectroscopic follow-up yet. Even if there is significant contamination from low redshift interlopers or variable objects, all our results still hold.
\item After reproducing the analysis from \cite{Ouchi2010}, we find that the main difference with our results is that we can more accurately include the brightest objects in our fit of the luminosity function and that we do a correction for our observational biases originating from the filter profile. We have additionally varied our EW criteria and used near-infrared data to exclude lower redshift interlopers, but find that this does not cause significant differences with the previous results for the ultra-deep surveys. For the deep COSMOS and shallower SA22 data, near-infrared data is however crucial. 
\item Comparing our $z=6.6$ data to lower redshift data at $z=5.7$, we confirm that there is evolution of the number density of the fainter LAEs. However, there is no evolution at the bright end and there is also evidence that the Ly$\alpha$ LF deviates from a Schechter function at the brightest luminosities.
\item Based on our toy-model, we argue that this differential evolution can be a sign of re-ionisation not being completed yet. Our model assumes that LAEs are only observed if they are in an ionised sphere which is large enough for the Ly$\alpha$ photons to redshift out of resonance wavelength, and that they are the only source of ionising radiation. Because of this, we preferentially observe the brightest LAEs, which have been able to ionise their own surrounding enough to be observable. Faint LAEs can only be observed if they are in the ionised spheres of more luminous LAEs, or when they are strongly clustered.
\item Additional wide field data at $z=5.7$ and $z=7.3$ will provide useful constraints on the evolution of the brightest LAEs around the re-ionisation epoch.
\item Finally, we use our results to make predictions for the upcoming Hyper Suprime Cam surveys at the Subaru telescope and also for upcoming very wide, cosmological surveys such as \emph{J-PAS} and \emph{Euclid}.
\end{itemize}

\section*{Acknowledgments}
We thank the anonymous referee for the comments and suggestions which have improved the quality of this work. We thank Masami Ouchi for his useful comments on an earlier version of this paper.
JM acknowledges the support of a Huygens PhD fellowship from Leiden University and is thankful for the hospitality of the Center for Astronomy and Astrophysics of the University of Lisbon where part of this research has been done. DS acknowledges financial support from the Netherlands Organisation for Scientific research (NWO) through a Veni fellowship, from FCT through a FCT Investigator Starting Grant and Start-up Grant (IF/01154/2012/CP0189/CT0010) and from FCT grant PEst-OE/FIS/UI2751/2014. HR acknowledges support from the ERC Advanced Investigator program NewClusters 321271. We acknowledge the award of ESO DDT time (294.A-5018) for providing the possibility of a timely publication of this work.

Based on observations with the Subaru Telescope (Program IDs: our observations: S14A-086; archival: S05B-027, S06A-025, S06B-010, S07A-013, S07B-008, S08B-008 and S09A-017) and the W.M. Keck Observatory. The Subaru Telescope is operated by the National Astronomical Observatory of Japan. The W.M. Keck Observatory is operated as a scientific partnership among the California Institute of Technology, the University of California and the National Aeronautics and Space Administration. Based on observations made with ESO Telescopes at the La Silla Paranal Observatory under programme ID 294.A-5018. Based on observations obtained with MegaPrime/Megacam, a joint project of CFHT and CEA/IRFU, at the Canada-France-Hawaii Telescope (CFHT) which is operated by the National Research Council (NRC) of Canada, the Institut National des Science de l'Univers of the Centre National de la Recherche Scientifique (CNRS) of France, and the University of Hawaii. This work is based in part on data products produced at Terapix available at the Canadian Astronomy Data Centre as part of the Canada-France-Hawaii Telescope Legacy Survey, a collaborative project of NRC and CNRS. Based on data products from observations made with ESO Telescopes at the La Silla Paranal Observatory under ESO programme ID 179.A-2005 and on data products produced by TERAPIX and the Cambridge Astronomy Survey Unit on behalf of the UltraVISTA consortium.

In addition to the CFHT-LS and COSMOS-UltraVISTA surveys, we are grateful for the excellent data-sets from the UKIRT-DXS, SXDF and S-COSMOS survey teams, without these legacy surveys, this research would have been impossible. We have benefited greatly from the public available programming language {\sc Python}, including the {\sc numpy, matplotlib, pyfits, scipy} and {\sc astropy} packages, the astronomical imaging tools {\sc SExtractor, Swarp} and {\sc Scamp} and the indispensable {\sc Topcat} analysis tool \citep{Topcat}.

\bibliographystyle{mn2e.bst}
\bibliography{bib_LAEevo.bib}

\appendix

\bsp

\label{lastpage}

\end{document}